# Post-Quantum Security: Origin, Fundamentals, and Adoption

Johanna Barzen[0000-0001-8397-7973] and Frank Leymann[0000-0002-9123-259X]

University of Stuttgart, IAAS, Universitätsstr. 38, 70569 Stuttgart, Germany
{firstname.lastname}@iaas.uni-stuttgart.de

**Abstract.** We first describe the relation between discrete logarithms and two well-known asymmetric security schemes, RSA and Elliptic Curve Cryptography. Next, we present the foundations of lattice-based cryptography which is the basis of schemes that are considered to be safe against attacks by quantum algorithms (as well as by classical algorithms). Then we describe two such quantum-safe algorithms (Kyber and Dilithium) in more detail. Finally, we give a very brief and selective overview of a few actions currently taken by governments and industry as well as standardization in this area. The article has a pedagogical character, not presenting any new research results. Especially it strives towards being self-contained, e.g. the required mathematical foundations to understand post-quantum cryptography are provided and examples are given. Thus, a reader interested in getting a first comprehensive overview on the subject doesn't have to consult text books or several research papers.

## 1. Introduction

Shor's algorithm [S97] for prime factorization is a major breakthrough in quantum computing: It allows to break RSA-based [RSA78] as well as elliptic curve-based cryptosystems exponentially faster than the best known classical algorithms. Thus, it is a severe threat to the established security infrastructure. But the quantum resources required to run Shor's algorithm (and thus to break today's security) have been considered at that time to be available in some decades only (if at all). This resulted in awareness of the problem itself but positioning it as a problem that must be dealt with in the far future. These days it seems likely that the required quantum resources are available in the next few years. Thus, today's security infrastructure is in jeopardy forcing policy makers and industry to urgently focus on this problem.

Prime factorization can be reduced to period finding of the module exponential function (section 3.1). Quantum computers are able to fast determine global properties of functions like their periods [NC16], which is exploited in Shor's algorithm by using the quantum Fourier transformation (section 3.2). Effectively, this way discrete logarithms are computed (section 2.5). Computing discrete logarithms also allows to compute private keys used in elliptic curve cryptography (section 2.6). Thus, a security scheme is needed that cannot be broken with discrete logarithms and that have not been cracked (yet).

Lattice-based cryptography is such a scheme. A lattice is the set of all integer combinations of linear independent vectors (section 4.1) or, equivalently, the set of all integer solutions of linear equation systems. While linear equations can be solved efficiently with classical methods, adding a small random error term to the equation system results in a hard problem (section 4.8) referred to as Learning with Errors. It is this problem that is the underpinning of algebraic lattice cryptography (chapter 5). Furthermore, instead of solving the Learning with Errors problem in vector spaces, modules over rings of residue classes of polynomials are used (section 5.3).

Based on these structures, an asymmetric encryption/decryption mechanism can be defined (section 5.4) as well as a means for signatures (section 5.5). These two mechanisms are recognized to be quantum-safe, i.e. neither classical algorithms nor quantum algorithms are known that can break these security schemes. Because of this, they are standardized as the basis for a future quantum-safe infrastructure (section 6.4).

The importance of becoming quantum-safe has been recognized by several government organizations, who initiated efforts to move today's infrastructure to a post-quantum secure one (section 6.3). This way, vital data like health data, military secrets etc should be kept confidential. To protect confidential data, companies must become active too, and the time needed for corresponding efforts can be estimated by Mosca's inequation (section 6.2). The corresponding transition is backed-up by open source software implementing the standardized post-quantum security algorithms (chapter 7).

### 1.1. Structure of the Article

The article is structured as follows: in the first part we give an overview on fundamental concepts and algorithms underlying today's security infrastructure. The second part sketches roughly a few actions of policy makers. While the first part (sections 2 to 5) contains timeless information, the much shorter second part (sections 6 and 7) is somehow volatile presenting information that might be outdated in a few years and that has to be digested by the reader with corresponding care.

## 2. Classical Asymmetric Cryptography

In this section we first remind a few definitions and facts from number theory related to modular arithmetic (see e.g. [MMS22]). Next, we provide the basics of RSA and elliptic curves which are the underpinnings of today's asymmetric cryptography (see [DK15] for details). Especially, their relation to discrete logarithms are presented.

## 2.1. Modular Arithmetic

A couple of prominent algorithms in cryptography like RSA (section 2.4) or Diffie-Hellman (section 2.6) are based on modular arithmetic. In this section we sketch its basics.

For $x \in \mathbb{R}$ let $\lfloor x \rfloor$ denote the greatest integer less than or equal to x (also called the *floor* of x). Then, for $a \in \mathbb{Z}$ and $n \in \mathbb{N}$ (i.e. especially $n \neq 0$), the *remainder* of $a$ divided by $n$ is given by

$$a - \left\lfloor \frac{a}{n} \right\rfloor n$$

and this remainder is denoted by "$a \mod n$" (pronounced "$a$ modulo $n$"). This is captured by the following

**Definition 1**: The function

$$\begin{aligned} \text{mod} : \mathbb{Z} \times \mathbb{N}^+ &\to \mathbb{N}_0 \\ (a, n) &\mapsto a - \left\lfloor \frac{a}{n} \right\rfloor n \quad =: a \text{ mod } n \end{aligned} \tag{1}$$

is called *modulo function*. □

The definition directly implies the following note:

**Note 1**: Let $a \in \mathbb{Z}$ and $n \in \mathbb{N}^+$. Then:

$$r = a \mod n \Leftrightarrow \exists\, k \in \mathbb{Z} : a = kn + r \wedge 0 \leq r < n \tag{2}$$ ∎

In case $a \mod n = b \mod n$, $a$ and $b$ give the same remainder when being divided by $n$. This is an important relation between the two numbers $a$ and $b$ that deserves a separate definition:

**Definition 2**: Let $a, b \in \mathbb{Z}$ and $n \in \mathbb{N}^+$. Then:

$$a \equiv b \pmod{n} :\Leftrightarrow a \mod n = b \mod n \tag{3}$$

In this case, $a$ and $b$ are called *congruent* modulo n. □

Congruence "≡" is an equivalence relation, i.e. it is reflexive, symmetric and transitive [MMS22]. Note that while "$a \equiv b \pmod{n}$" is a Boolean statement (i.e. it is true or false for given $a, b, n$), "$a \mod n$" is a natural number (i.e. the remainder of $a$ divided by $n$).

**Definition 3**: Let $m \in \mathbb{Z}$ and $n \in \mathbb{N}^+$. The notation $n \mid m$ is defined as follows:

$$n \mid m :\Leftrightarrow \exists\, k \in \mathbb{Z} : m = kn \tag{4}$$

$n$ is called a *divisor* of $m$ (aka $n$ *divides* $m$ aka $n$ is a *factor* of $m$). The notation $n \nmid m$ means that $n$ does not divide $m$. In case $n \neq 1$ and $n \neq m$, $n$ is called a *proper divisor* of $m$. □

The next note follows immediately from the definitions before :

**Note 2**:

$$a \equiv b \ (\text{mod } n) \Leftrightarrow n \,|\, (a - b) \qquad (5) \quad ∎$$

Proving the congruence of two numbers (especially of large numbers) can be very tedious. The following often simplifies related computations:

**Principle (Reduce First, Then Compute)**: Assume that $(a_1 a_2 + a_3 a_4) \text{ mod } n$ has to be computed, and let $a_i \equiv b_i \text{ mod } n$ for $1 \leq i \leq 4$. Then:

$$(a_1 a_2 + a_3 a_4) \text{ mod } n = (b_1 b_2 + b_3 b_4) \text{ mod } n$$

The proof follows directly from the definitions.

This principle allows to reduce an arithmetic expression that might be cumbersome to compute (in our case $a_1 a_2 + a_3 a_4$) first and then use an arithmetic expression that might be easier to compute (in our case $b_1 b_2 + b_3 b_4$) to determine the remainder divided by $n$.

Example [MMS22]: $(39 \cdot 99 + 95 \cdot 64) \text{ mod } 19$ should be computed. Now, it is $39 \equiv 1 \text{ mod } 19$, $99 \equiv 4 \text{ mod } 19$, $95 \equiv 0 \text{ mod } 19$, and $64 \equiv 7 \text{ mod } 19$. Thus, we can substitute 39 by 1, 99 by 4, 95 by 0, and 67 by 7 and result in

$$(39 \cdot 99 + 95 \cdot 64) \equiv (1 \cdot 4 + 0 \cdot 7) \equiv 4 \text{ mod } 19 = 4$$

which is much simpler then computing $(39 \cdot 99 + 95 \cdot 64)$ first and then finding the remainder divided by 19.

Another example is computing $127^4 \text{ mod } 5$. It is $127 \equiv 2 \text{ mod } 5$. Thus, it is

$$127^4 \equiv 127 \cdot 127 \cdot 127 \cdot 127 \equiv 2 \cdot 2 \cdot 2 \cdot 2 \equiv 16 \text{ mod } 5 = 1$$

In our final example, we want to compute $129537 \text{ mod } 9$. This can be done by using a decimal polynomial representation of 129537 and observing $10 \equiv 1 \text{ mod } 9$:

$$\begin{aligned}
129537 &= 1 \cdot 10^5 + 2 \cdot 10^4 + 9 \cdot 10^3 + 5 \cdot 10^2 + 3 \cdot 10^1 + 7 \cdot 10^0 \\
&\equiv 1 \cdot 1^5 + 2 \cdot 1^4 + 9 \cdot 1^3 + 5 \cdot 1^2 + 3 \cdot 1^1 + 7 \cdot 1^0 \\
&\equiv 1 + 2 + 9 + 5 + 3 + 7 \\
&\equiv 27 \\
&\equiv 0 \text{ mod } 9 \\
&= 0
\end{aligned}$$

Thus, the principle "reduce first then compute" is a fundamental principle in *modular arithmetic* to make computations easier.

## 2.2. Diophantine Equations

Several algorithms in cryptography require to solve *linear Diophantine equations*, i.e. equations of the form

$$ax + by = c \qquad (6)$$

with given $a, b, c \in \mathbb{N}$ and where $x, y \in \mathbb{Z}$ (i.e. the solution of the equation) have to be determined. Deciding whether or not a linear Diophantine equation has a solution requires to determine greatest common divisors.

**Definition 4**: $n \in \mathbb{N}^+$ is called a *common divisor* of $a, b \in \mathbb{Z}$ iff both, $n \mid a$ and $n \mid b$. The *greatest common divisor* of $a$ and $b$ is denoted by $\gcd(a, b)$. $a$ and $b$ are called *coprime* iff $\gcd(a, b) = 1$. □

The next Theorem provides a criterion for the solvability of a linear Diophantine equation.

**Theorem 1**: Let $a, b, c \in \mathbb{N}$. Then: $\left(\exists\, x, z \in \mathbb{Z} : a x + b y = c\right) \Leftrightarrow \gcd(a, b) \mid c$

Proof: [MMS22] ∎

The famous *Euclidian Algorithm* that is the basis for computing solutions of linear Diophantine equations is given by

**Theorem 2**: Let $a, b \in \mathbb{N}$. Define $r_0 := a$, $r_1 := b$ and $r_{i+1} := r_{i-1} \bmod r_i$ for $i \geq 1$. Then there exists an $N \in \mathbb{N}$ such that $r_N = 0$ and $r_{N-1} = \gcd(a, b)$.

Proof: [MMS22] ∎

The (extended) Euclidian Algorithm is often applied by using a certain schema when writing down the computation steps (see [MMS22] section 3.2 for the details).

As an example, we apply this schema to compute the greatest common divisor gcd(63,17) of the numbers 63 and 27 (the numbering of the equations in needed in the next example):

$$\begin{aligned}
63 &= 3 \cdot 17 + 12 && \text{(d)}\\
17 &= 1 \cdot 12 + 5 && \text{(c)}\\
12 &= 2 \cdot 5 + 2 && \text{(b)}\\
5 &= 2 \cdot 2 + 1 && \text{(a)}\\
2 &= 2 \cdot 1 + 0
\end{aligned}$$

Thus, $\gcd(63,17) = 1$, i.e. 63 and 17 are coprime.

Within Theorem 1, the direction "⇐" is called *Bézout's Identity*. Because of its importance it is stated as a separate lemma.

**Lemma 1**: Let $a, b \in \mathbb{N}$ and let $g = \gcd(a, b)$. There exist $x, y \in \mathbb{Z}$ such that

$$g = a x + b y \qquad (7)$$ ∎

The factors x and y are called *Bézout coefficients* for $(a, b)$ and can be determined based on the Euclidian algorithm. For this purpose, the equations of the computations performed according to the Euclidian algorithm are applied "from bottom to top" beginning with the last equation, i.e. the one containing the gcd (see [MMS22] section 3.2 for the details).

As an example, we use the computation of gcd(63,17) before: Using equation (a) results in

$$1 \overset{(a)}{=} 5 - 2 \cdot 2$$

Base on equation (b), it is $2 \overset{(b)}{=} 12 - 2 \cdot 5$, which is substituted in the equation before (and so on):

$$\begin{aligned} 1 &\overset{(a)}{=} 5 - 2 \cdot 2 \\ &\overset{(b)}{=} 5 - 2 \cdot (12 - 2 \cdot 5) = 5 \cdot 5 - 2 \cdot 12 \\ &\overset{(c)}{=} 5 \cdot (17 - 1 \cdot 12) - 2 \cdot 12 = 5 \cdot 17 - 7 \cdot 12 \\ &\overset{(d)}{=} 5 \cdot 17 - 7 \cdot (63 - 3 \cdot 17) = 5 \cdot 17 - 7 \cdot 63 + 21 \cdot 17 \\ &= 26 \cdot 17 - 7 \cdot 63 \end{aligned}$$

Thus, the Bézout coefficients for (63,17) are (−7,26).

This proceeding can be used to solve an equation important for RSA (section 2.4 (5)) following the proof of

**Note 3**: Let $a, m \in \mathbb{N}$ be coprime, i.e. $\gcd(a, m) = 1$. Then, there exists an $x \in \mathbb{Z}$ such that $ax \equiv 1 \bmod m$.

Proof: Because of lemma 1, there exist $x, y \in \mathbb{Z}$ such that $1 = ax + my$. This implies that $ax - 1 = -my$. According to definition 3, this means $m \mid (ax - 1)$. With note 2 we get $ax \equiv 1 \bmod m$. ∎

## 2.3. Factorization

One of the key algorithms in cryptography is based on factorization of numbers (see section 2.4). Because of this, we remind the basics of factorization. Factorization means to determine the composition of an integer of prime numbers.

**Definition 5**: A number $p \in \mathbb{N}$ with no proper divisor is called a *prime number*. The set of all prime numbers is denoted by $\mathbb{P}$. □

Every natural number can be represented as a product of prime numbers, and this representation is unique up to order of the prime numbers in the product: this is the famous *Fundamental Theorem of Arithmetic*. Determining the prime numbers of this representation of a number is referred to as its *factorization*.

**Theorem 3**: Let $n \in \mathbb{N}$ with $n > 1$. There exist $p_1, \ldots, p_k \in \mathbb{P}$ ($p_i \neq p_j$ for $i \neq j$) and $n_1, \ldots, n_k \in \mathbb{N}$ such that

$$n = p_1^{n_1} \cdot p_2^{n_2} \cdot \ldots \cdot p_k^{n_k} \tag{8}$$ ∎

There are many known classical algorithms to compute the factorization of a given number. None of these known algorithms is efficient in the sense that it can compute the factorization in polynomial time. But is has not been proven that a polynomial algorithm cannot exist, nor has it been proven that such an algorithm does exist. Thus, it is open whether or not factorization can be done efficiently.

In what follows, we will make use of the *Lemma of Euclid*.

**Lemma 2**: Let $p \in \mathbb{P}$ and $a, b \in \mathbb{N}$. If $p \mid ab$ then $p \mid a$ or $p \mid b$. ∎

Assume $n \in \mathbb{N}$ with $n \mid ab$ and let $n = p_1^{n1} \cdot p_2^{n2} \cdot \ldots \cdot p_k^{nk}$ the factorization of *n*. Because $ab = mn = mp_1^{n1} \cdot p_2^{n2} \cdot \ldots \cdot p_k^{nk} = (mp_1^{n1} \cdot \ldots \cdot p_i^{n_i - 1} \cdot \ldots \cdot p_k^{nk}) \cdot p_i$ for a suitable $m \in \mathbb{N}$, it is $p_i \mid ab$ for each $1 \leq i \leq k$. According to Euclid's lemma before, it is $p_i \mid a$ or $p_i \mid b$ for $1 \leq i \leq k$. Thus, each such $p_i$ is a common divisor of *n* as well as of *a* or *b*. This proves the following corollary:

**Corollary 1**: Let $n \in \mathbb{N}$ with $n \mid ab$. Then: $\exists\, m \in \mathbb{N} : m \mid n \wedge (m \mid a \vee m \mid b)$. Especially, if *n* is not prime, *m* is a proper divisor of *n*. ∎

Note, that in general $n \mid ab$ does not imply that $n \mid a$ or $n \mid b$. For example: $6 \mid 12 = 3 \cdot 4$ but neither $6 \mid 3$ nor $6 \mid 4$; but of course, 3 is common divisor of 6 and 3, and 2 is a common divisor of 6 and 4.

## 2.4. RSA

RSA is a public-key cryptosystem which as based on the assumed hardness of factorization of natural numbers. Public and private keys are determined as follows:

1. Choose two large prime numbers $p, q \in \mathbb{P}$
2. Compute $n = pq$
3. Compute $\varphi(n) = (p-1)(q-1)$
4. Choose a small $g \in \mathbb{N}$, with $\varphi(n)$ and $g$ being coprime (i.e. $\gcd(\varphi(n), g) = 1$)
5. Determine the solution $d$ of the equation $dg \equiv 1 \bmod \varphi(n)$ (see Note 3)
6. Destroy $p, q, \varphi(n)$

(*d, n*) is the *public key*, and *g* is the *private key*.

The following examples are from [RSA78]. Chose the prime number $p = 47$ and $q = 59$ (which are obviously not "large" but fine for the example). Then $n = p \cdot q = 2773$ and $\varphi(n) = 46 \cdot 58 = 2668$. The private key is chosen as $g = 157$ which is valid because $\gcd(2668{,}157) = 1$, i.e. $g$ and $\varphi(n)$ are coprime. Next, the equation $d \cdot 157 \equiv 1 \bmod 2668$ is solved according to Note 3 by computing the Bézout coefficients for (2668,157) as exemplified in the example after Lemma 1: the solution is $d = 17$. Thus, (17,2773) is the public key.

A message $\mu$ is encrypted as follows:

1. $\mu$ is mapped to a natural number $m$
   a. E.g. each letter is substituted by a two-digit number like Blank=00, A=01, B=02 etc., and the numbers corresponding to the letters of the message are concatenated.
   b. The resulting number $m$ must be less than n. If $m \geq n$, the digits of $m$ are split into blocks $m_1, \ldots, m_k$ such that $m_i < n$.
2. Each block is encrypted as $\widehat{m}_i = m_i^d \bmod n$
3. The encrypted message is the concatenation $\widehat{m} = \widehat{m}_1 \cdots \widehat{m}_k$

Example: Let $\mu$ = ITS ALL GREEK TO ME. By mapping this message to a natural number as sketched in (1a) and considering (1b) we get m = 0920 1900 0112 1200 0718 0505 1100 2015 0013 0500, i.e. m has been split into blocks of digits the corresponding numbers are less than $n = 2773$. The first block $m_1$ corresponds to the number 920, thus, according to (2) the encryption of this block $m_1$ results in $\widehat{m}_1 = 920^{17} \bmod 2773 = 948$; the second block corresponds to 1900, i.e. $\widehat{m}_2 = 1900^{17} \bmod 2773 = 2342$ etc. Finally, the encrypted message is according to (3) the concatenation of the encrypted blocks: $\widehat{m} = 0948\ 2342\ldots$

Decryption is achieve by

4. decrypting each block $\widehat{m}_i$ of the encrypted message $\widehat{m}$ separately
5. by computing $m_i = (\widehat{m}_i)^g \bmod n$
6. and concatenating the decrypted block resulting in the decrypted message $m = m_1 \cdots m_k$

Example: In the encrypted message $\widehat{m}$ from above, it is $\widehat{m}_1 = 948$, i.e. according to (5) it is $m_1 = (\widehat{m}_1)^{157} \bmod 2773 = 984^{157} \bmod 2773 = 920$. With (1a) 920 corresponds to IT (i.e. 09=I, 20=T) and so on. Concatenating the decrypted block results in the original message.

Cracking a private key means to compute $g$. While $g$ is secrete, i.e. unknown, $d$ as part of the public key is known. To compute $g$, the equation $dg \equiv 1 \bmod \varphi(n)$ must be solved which requires in addition to $d$ to know $\varphi(n)$. This, in turn, means that $\varphi(n) = (p-1) \cdot (q-1)$ must be known. Because the RSA procedure is well-known, it is further known that $n$ is the product of two prime numbers. Consequently, $n$ must "just" be factorized to get $\varphi(n)$ and then solve $dg \equiv 1 \bmod \varphi(n)$. Since factorization is assumed to be hard, RSA is considered hard to be cracked, i.e. RSA is considered to be secure (ignoring Shor's algorithm for now).

### 2.5. Modular Exponential Function and Discrete Logarithm

Factorization is closely related to the following function and its inverse, respectively:

**Definition 6**: For a given $n \in \mathbb{N}$ choose an arbitrary number $a \in \mathbb{N}$ with $0 < a < n$. The function

$$\exp_a : \mathbb{N}_0 \to \mathbb{N}_0 \text{ with } \exp_a(x) = a^x \bmod n \tag{9}$$

is called *modular exponential function* with basis $a$ (and modulo $n$). □

As an example, consider the first part of the value table of the modular exponential function with $n = 5$ and basis 2, i.e. :

| x | 0 | 1 | 2 | 3 | 4 | 5 | 6 | 7 | … |
|---|---|---|---|---|---|---|---|---|---|
| $\exp_2(x)$ | 1 | 2 | 4 | 3 | 1 | 2 | 4 | 3 | … |

As can be seen, it is $\exp_2(x+4)=\exp_2(x)$. I.e. the function $\exp_2(\cdot)$ with $n=5$ seems to be periodic with period 4. These terms are defined next.

**Definition 7**: A function $f:\mathbb{N}_0\to\mathbb{N}_0$ is called *periodic* $:\Leftrightarrow\ \exists\, t\in\mathbb{N}\ \forall\, x\in\mathbb{N}_0:$ $f(x+t)=f(x)$. The smallest such number $t$ is called the *period* of *f*. □

The following is a well-known fact (a corollary of Fermat's Little Theorem) about the periodicity of the modular exponential function:

**Lemma 3**: The modular exponential function $\exp_a(x)=a^x \bmod n$ is periodic if and only if $\gcd(n,a)=1$. ∎

Let $\gcd(n,a)=1$, thus, $\exp_a(x)$ is periodic with period $p\in\mathbb{N}$. It is then $\exp_a(0)=\exp_a(0+p)=\exp_a(p)$ which in turn is equivalent to $a^p \bmod n = a^0 \bmod n = 1 \bmod n$. By Definition 2, this is equivalent to $a^p\equiv 1 \pmod n$, and by Note 2 we know that $n\,|\,(a^p-1)$, i.e. $n$ is a divisor of $a^p-1$. This is summarized by

**Note 3**: Let $a,n\in\mathbb{N}$ be coprime. $\exp_a(x)=a^x \bmod n$ has period $p\in\mathbb{N}$ if and only if $a^p\equiv 1 \pmod n$, i.e. $n\,|\,(a^p-1)$. ∎

Let $n\,|\,(a^p-1)$ and assume $p\in\mathbb{N}$ is even. Then, $n\,|\,(a^{p/2}-1)(a^{p/2}+1)$, and according to Corollary 1, there is an $m\in\mathbb{N}$ with $m\,|\,n$ and $m\,|\,(a^{p/2}-1)$ or $m\,|\,(a^{p/2}+1)$. Thus, $n$ and $(a^{p/2}-1)$ have a common divisor, or $n$ and $(a^{p/2}+1)$ have a common divisor. This implies that $\gcd((a^{p/2}-1),n)$ or $\gcd((a^{p/2}+1),n)$ is a divisor of $n$.

This means, if the period of $\exp_a$, $0<a<n$, can be determined, and the period is even, a divisor of $n$ can easily be computed. And by iteration, the factorization of $n$ results. Before discussing the frequency of this situation, we give the following

**Definition 8**: The smallest number $y\in\mathbb{N}$ with $a^y\equiv x \bmod n$ (and $0<a<n$) defines a map

$$\log_a:\mathbb{N}_0\to\mathbb{N}_0 \text{ with } x\mapsto y \tag{10}$$

called *discrete logarithm* (of x) with basis $a$ (and modulo $n$). □

As an example, it is $2^4=16\equiv 5 \bmod 11$, with implies $\log_2 5=4$ (modulo 11). The usual equations like $\exp_a(\log_a z)=z$ and $\log_a(\exp_a y)=y$ apply the discrete logarithm and modular exponential function.

Computing the period $p$ of $\exp_a$ means to determine $p\in\mathbb{N}_0$ with $a^p\equiv 1 \pmod n$ (see Note 3), i.e. to compute the discrete logarithm $p=\log_a 1$. This means that factorization - and, thus, cracking RSA - can be reduced to compute discrete logarithms! But computing discrete logarithms is assumed to be classically hard, just like factorization. Similarly, cracking cryptography based on elliptic curves can be reduced to computing discrete logarithms (see section 2.6).

Remember that the assumption we made in order to determine the period $p\in\mathbb{N}$ of $\exp_a$, $0<a<n$, is that $p$ is even. Under this assumption, $\gcd((a^{p/2}-1),n)$ or $\gcd((a^{p/2}+1),n)$ is a divisor of $n$. First of all, the question is how often the period is even. Next, the question is whether $\gcd((a^{p/2}-1),n)$ or $\gcd((a^{p/2}+1),n)$ are proper divisors of $n$. The first observation is:

**Note 4**: $\gcd((a^{p/2}-1),\, n) \neq n$.

Proof: Let $\gcd((a^{p/2}-1),\, n) = x$. Then, we find $y, z \in \mathbb{N}$ with $zx = a^{p/2}-1$ and $yx = n$. Thus, $a^{p/2}-1 = zx = z \cdot n/y = n \cdot z/y$. Assume $x = n$, thus, $y = 1$. This means that $a^{p/2}-1 = nz \Rightarrow a^{p/2} = nz + 1 \Rightarrow a^{p/2} \equiv 1 \bmod n$. According to Note 3 this means that the period of $\exp_a$ is $p/2$ - which is a contradiction. This proves the claim. ∎

It may still be the case that $\gcd((a^{p/2}-1),\, n) = 1$, i.e. that $\gcd((a^{p/2}-1),\, n)$ is not a proper divisor of $n$. In case $\gcd((a^{p/2}-1),\, n) = 1$, Corollary 1 implies that $\gcd((a^{p/2}+1),n) \neq 1$. Still, it may be that $\gcd((a^{p/2}+1),n) = n$, and, thus, $\gcd((a^{p/2}+1),n)$ may not be a proper divisor of n.

Consequently, the bad situation $\gcd((a^{p/2}-1),\, n) = 1$ and $\gcd((a^{p/2}+1),n) = n$ may occur. But according to ([NC16], Theorem A4.13) the following is known:

**Note 5**: For more than half of the numbers $0 < a < n$ with $\gcd(n,a) = 1$, the period $p$ of $\exp_a$ is even, and $\gcd((a^{p/2}-1),\, n) \neq 1$ or $\gcd((a^{p/2}+1),n) \neq n$. ∎

Thus, after a couple of attempts a number $a$ will be found such that $\exp_a$ has an even period and that $\gcd((a^{p/2}-1),\, n)$ or $\gcd((a^{p/2}+1),n)$ is a proper divisor of $n$.

How the period of $\exp_a$ can be determined efficiently is subject of section 3.

### 2.6. Elliptic Curve Cryptography

Elliptic curve cryptography can also be broken by efficient computations of discrete logarithms on quantum computers. In this section, we briefly introduce the underpinnings of elliptic curve cryptography and sketch its relationship to discrete logarithms. For more details about elliptic curves as well as the proofs of the facts given in this section, see [R18].

Note, that an elliptic curve is not an ellipsis. But the name results from computing the arc length of an ellipsis by means of integrals. With proper parametrization the integrands of the corresponding integrals contain polynomial functions as used in the definition of the following set *C*:

**Definition 9**: Let $C := \left\{(x,y) \in \mathbb{R}^2 \mid y^2 = x^3 + ax + b\right\} \cup \{\mathcal{N}\}$ and $a, b \in \mathbb{R}$ with $4a^3 + 27b^2 \neq 0$. Then *C* is called an *elliptic curve*. □

The reason why $\mathcal{N}$ ("the infinitely far point") is added will get clear soon. The condition $4a^3 + 27b^2 \neq 0$ excludes repeated roots that would result in singularities like intersections or cusps of the curve. Figure 1 depicts three elliptic curves to give an impression about their possible shapes.

| $y^2 = x^3 - 2x + 1$ | $y^2 = x^3 - 2x + 2$ | $y^2 = x^3 + x + 1$ |
|---|---|---|

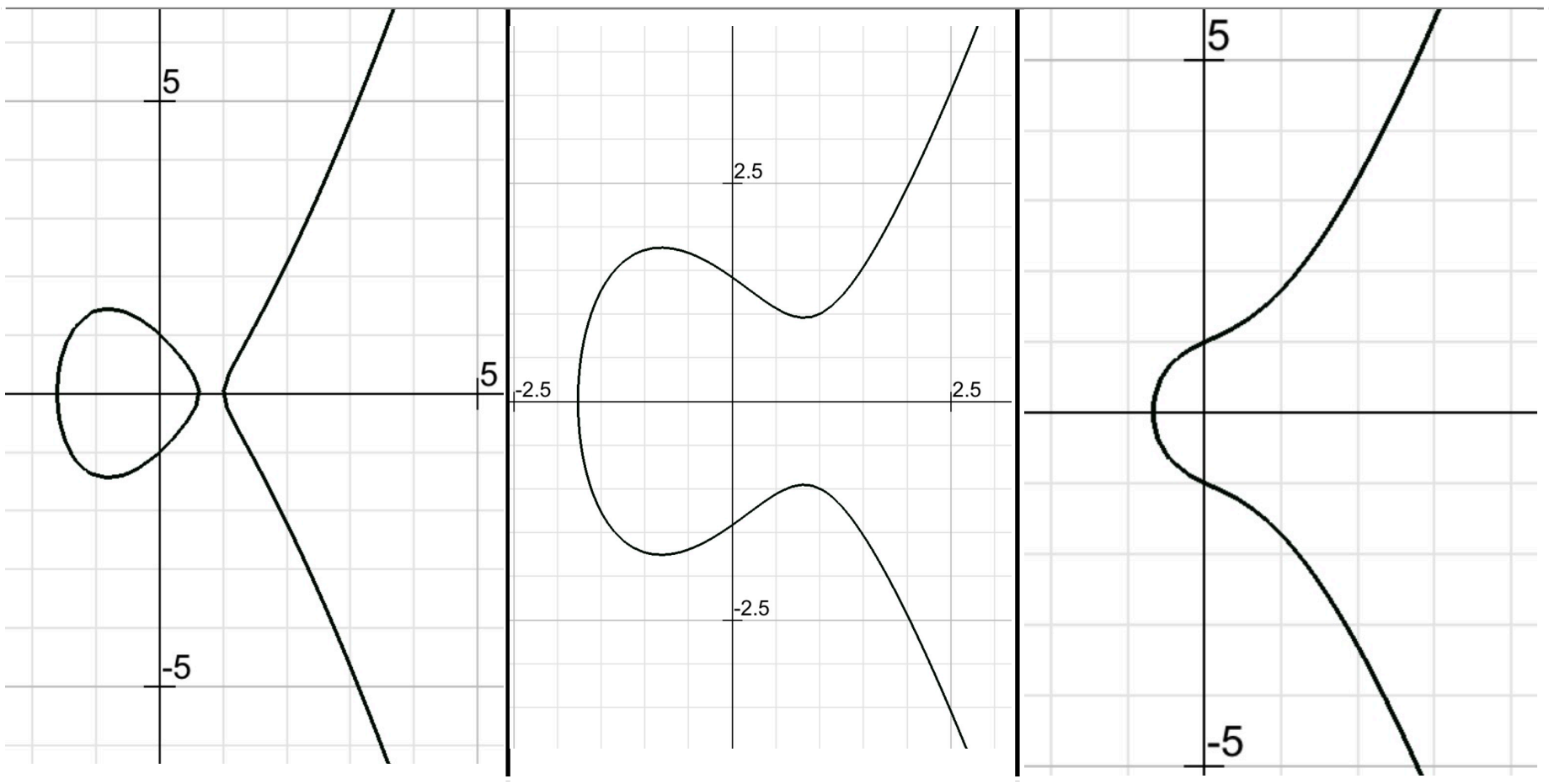

**Fig. 1**. Sample Shapes of Elliptic Curves

An elliptic curve $C$ is obviously symmetrical to the x-axis, i.e. for each point $P \in C$ on the curve, its reflection across the x-axis is again a point on the curve, which is denoted by $P^{-1} \in C$. This allows to define a multiplication $\circ : C \times C \to C$ on a curve (see Figure 2) as follows.

1. Let $P \neq Q \in C$ be two distinct points on the curve $C$ and let the line $\overline{PQ}$ intersect the curve $C$ in the point $Z$: $\overline{PQ} \cap C = Z$. The reflection $Z^{-1}$ of $Z$ across the x-axis is defined as the product of $P$ and $Q$: $P \circ Q := Z^{-1}$
2. Let $R \in C$ be a point, let $g_R$ be the tangent to $C$ at $R$, and let $W$ be the intersection of the tangent with $C$: $g_R \cap C = W$. The reflection $W^{-1}$ of $W$ across the x-axis is defined as the product of $R$ with itself: $R^2 := W^{-1}$.
3. Let $S \in C$ be a point whose tangent does not intersect $C$. For this purpose, the curve $C$ is extended by a point at infinity $\mathcal{N}$, and the product of $S$ with itself is this point: $S^2 := \mathcal{N}$.
4. Let $T \in C$ and $T^{-1} \in C$ its reflection across the x-axis. The line connecting $T$ and $T^{-1}$ is parallel to the y-axis and has no intersection with the curve. The product of $T$ and $T^{-1}$ is defined as the infinitely far point $\mathcal{N}$: $T \circ T^{-1} := \mathcal{N}$.

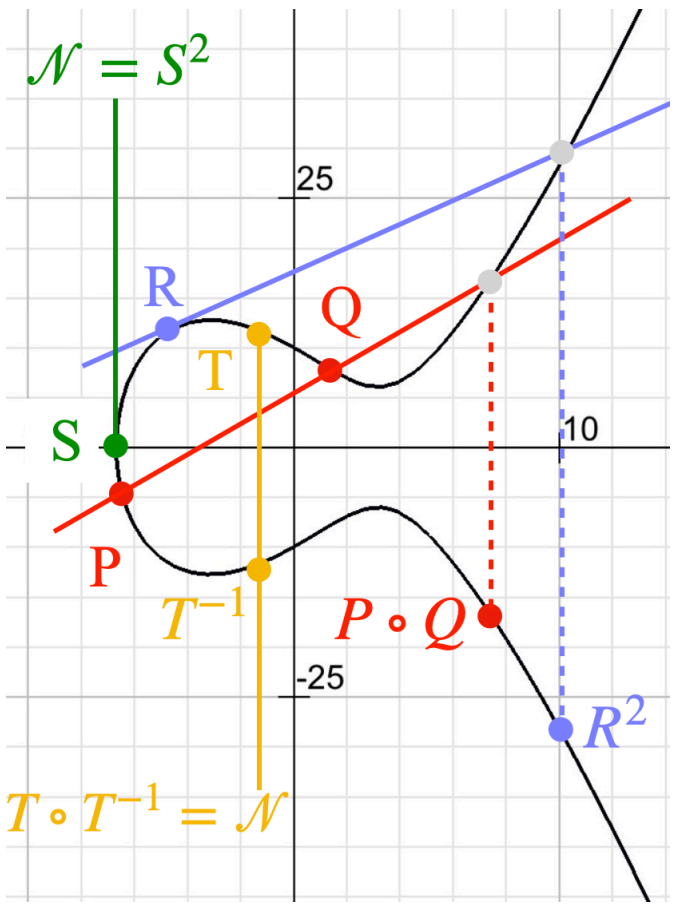

**Fig. 2**. Operations on an Elliptic Curve

With respect to this operation $\circ$ , the reflection $P^{-1}$ of a point $P \in C$ across the x-axis is the inverse element of $P$, and the infinitely far point $\mathcal{N}$ is revealed as the neutral element of the operation. The use of elliptic curves in cryptography is based on this operation and the corresponding

**Theorem 3**: Let $C$ be an elliptic curve. Then $(C, \circ)$ is a commutative group with the neutral element $\mathcal{N}$. ∎

For use of an elliptic curve in a cryptosystem the parameters $a, b \in \mathbb{R}$ of the curve are published as well as a point $G \in C$ called the *generator*. Each participant has a *private key* $s \in \mathbb{N}$ and $G^s =: P \in C$ as the corresponding *public key*. The following example shows how elliptic curve encryption works:

A message $\mu$ is encrypted as follows:

1. $\mu$ is mapped to a point $M \in C$, e.g.:
   a. $\mu$ is considered as a bit string $m \in \mathbb{N}$
   b. Compute $y$ with $y^2 = m^3 + am + b$
   c. Then $M = (m, y) \in C$
2. Chose a random $k \in \mathbb{N}$
3. Compute $E_1 = G^k$ and $E_2 = M \circ P^k$
   - $G$ is the published generator, $s$ is private key to compute $P = G^s$
4. The encrypted message is $\chi(m) = (E_1, E_2)$

An encrypted message $\chi(m)$ is decrypted as follows:

5. Compute $M = E_2 \circ (E_1)^{-s}$
   - It is $(E_1)^s = \left(G^k\right)^s = \left(G^s\right)^k = P^k$
   - Thus, it is $E_2 \circ (E_1)^{-s} = M \circ P^k \cdot (E_1)^{-s} = M \circ P^k \circ P^{-k} = M$

6. The decrypted message is $m = \pi_1(M)$

Note, that in practice elliptic curve cryptography (ECC) is is not used for encryption of persistent data but for key exchange and signatures. For example, elliptic curve key exchange (a.k.a. ECC Diffie-Hellman) works as follows:

1. As before, let $G \in C$ be the generator of the curve
2. Alice has $G^{s_A} =: P_A \in C$ as public key, with $s_A \in \mathbb{N}$ as private key
3. Bob has $G^{s_B} =: P_B \in C$ as public key, with $s_B \in \mathbb{N}$ as private key
4. Alice and Bob exchange $P_A, P_B$
5. Alice computes $K_A = P_B^{s_A}$
6. Bob computes $K_B = P_A^{s_B}$
   - It is $K_A = P_B^{s_A} = \left(G^{s_B}\right)^{s_A} = \left(G^{s_A}\right)^{s_B} = P_A^{s_B} = K_B$
   - Thus, Alice and Bob share the same secrete key $K \left( = K_A = K_B\right)$
7. Alice and Bob can now use the shared secrete key for symmetric encryption

A public key is of the form $P = G^s$ with the publicly known generator $G \in C$. This means that determining the private key $s$ ("cracking elliptic curve cryptography") means to compute $s = \log_G P$ where the exponential $\exp_G$ is computed based on "∘" as multiplication. Consequently, cracking elliptic curve cryptography is also reduced to compute discrete logarithms.

Note, that elliptic curves relevant for cryptography are curves whose coefficients are from a finite field $\mathbb{F}_p$ instead of coefficients from $\mathbb{R}$. The principles sketched before remain valid, i.e. this complication is irrelevant for a first treatment and understanding.

# 3. Shor's Algorithm

The algorithm of Shor [S97] for factorization is a hybrid algorithm, i.e. it consists of classical computations as well as quantum computations. The same paper describes a modification of the algorithm for computing discrete logarithms. Overall, both algorithms provide an exponential speedup compared with the best know classical algorithms for solving these two problems. In this section, we very briefly sketch how numbers are factorized with Shor's algorithm.

## 3.1. The Outer Classical Part

The classical computations can be split into an "outer" classical part that surrounds both, the quantum part (section 3.2) as well as an "inner" classical part (section 3.3). The outer part computes greatest common divisors and decides whether to iterate the overall algorithm or to stop with a divisor of the number to be factorized.

Let $n \in \mathbb{N}$ be the number to be factorized ($n$ can be assumed to be odd because otherwise its factor "2" is immediately known). Randomly, a number $a \in \mathbb{N}$, $1 < a < n$ is chosen. In case $\gcd(a, n) \neq 1$ a divisor of $n$ has been found and the algorithm stops. The algorithm continues with $\gcd(a, n) = 1$, and according to Lemma 3 the function $\exp_a(x) = a^x \bmod n$ is periodic.

Next, the quantum part (section 3.2) and the inner classical part (section 3.3) are used to determine the period of $\exp_a(x) = a^x \bmod n$. If it turns out that $p$ is odd, the algorithm starts over again with a different $1 < a < n$ because we need $p$ even to apply the representation $a^p + 1 = (a^{p/2} - 1)(a^{p/2} + 1)$. Then, according to Note 4, after a couple of attempts a number $a$ will be found such that $p$ is even, and that $\gcd(a^{p/2} - 1, n)$ or $\gcd((a^{p/2} + 1), n)$ is a proper divisor of $n$: the algorithm stops with a divisor.

### 3.2. The Quantum Part

The details of Shor's algorithm especially its quantum part can be found in most text books on quantum computing like [NC16]. Here, we briefly sketch the key facts (ignoring some details to simplify the treatise).

First, the integers $0 \leq x \leq N$ (with $N = 2^m$ and $n^2 \leq 2^m < 2n^2$) are prepared in a two-part quantum register by applying a Hadamard transformation:

$$H^{\otimes m} \otimes I(|0\rangle|0\rangle) = \frac{1}{\sqrt{N}} \sum_{x=0}^{N-1} |x\rangle|0\rangle \tag{11}$$

Next, an oracle is used to compute the value table of the modular exponential function of these numbers:

$$|\alpha\rangle|\beta\rangle = \frac{1}{\sqrt{N}} \sum_{x=0}^{N-1} |x\rangle|a^x \bmod n\rangle \tag{12}$$

Note the similarity with the value table after Definition 6; as discussed there, based on a value table of a function its periodicity can be determined “by inspection”. Such an inspection can be done by applying a quantum Fourier transform on the $|\alpha\rangle$-part of the quantum register, and a succeeding measurement of the $|\beta\rangle$-part. The resulting state of the $|\alpha\rangle$-part is :

$$\left( \frac{1}{\sqrt{NA}} \sum_{j=0}^{A-1} \exp\left(\frac{2\pi i}{N} jpy\right) \right) |y\rangle \tag{13}$$

The quantum part ends with measuring the $|\alpha\rangle$-part of the register resulting in the value $y$.

### 3.3. The Inner Classical Part

An extensive explanation of the inner classical part is given in [BL22].

Note that the period $p$ is already reflected in the amplitude of the state $|y\rangle$ reached after the Fourier transform (equation (13)): how can it be derived from this equation? According to the Born rule of quantum physics, the probability $\text{Prob}(y)$ to measure a particular $y$ is the squared modulus of the amplitude of the associated state $|y\rangle$, i.e. based on equation (13):

$$\text{Prob}(y) = \frac{1}{NA}\left|\sum_{j=0}^{A-1}\exp\left(\frac{2\pi i}{N}jpy\right)\right|^2 \tag{14}$$

The sum in equation (14) is a geometric sum, thus, the well-known formula for computing the value of such a sum can be used, and with $q := \exp\left(\frac{2\pi i}{N}py\right)$ we get

$$\text{Prob}(y) = \frac{1}{NA}\left|\frac{1-q^A}{1-q}\right|^2 \tag{15}$$

A detailed analysis of equation (15) reveals (see [BL22], section 3, for the details) that "with high probability" a number $k$ with $0 < k < p - 1$ can be found that

$$\left|\frac{y}{N} - \frac{k}{p}\right| \leq \frac{1}{2p^2} \tag{16}$$

Equation (16) is the pre-condition to apply Legendre's Theorem on continued fractions [BL22] which states that in this situation, $\frac{k}{p}$ is a convergent of $\frac{y}{N}$. Note that $y$ is known because it has been measured, and $N$ has been chosen at the beginning of the quantum part, i.e. $\frac{y}{N} \in \mathbb{Q}$ is known. Thus, the continued fraction representation $[a_0; a_1, \ldots, a_l]$ of $\frac{y}{N}$ can be derived, and one of the convergents $[a_0; a_1, \ldots, a_u] = \frac{g_u}{h_u}$ with $1 \leq u \leq l$ will be a "very good" approximation of $\frac{k}{p}$, i.e. the corresponding denominator $h_u$ will be a candidate for the period $p$ (note, that the actual value of $k$ is not needed because we are only interested in the denominator $p$). Whether or not this candidate $h_u$ is really the period or not must be checked explicitly.

Consequently, the inner classical part basically consists of a continued fraction analysis of $\frac{y}{N}$ that produces the period $p$ "with high probability". In case a period is not found, the overall algorithm starts over again with a different $1 < a < n$.

### 3.4. Quantum Ressource Requirements

Shor's algorithm can efficiently break cryptosystems that are based on factorization or elliptic curves. While its classical parts can be performed on any classical computer today, its quantum part assumes a quantum computer that has error corrected qubits and operations without errors.

Different authors discuss implementation variants of the quantum part of the factorization algorithm and state the number of logical qubits (i.e. error corrected qubits) and faithful operations (see [GE21] for an overview). As a very rough average, for RSA2048 the number of logical qubits needed is about $10^4$ and the number of operations is about $10^{11}$ (and the operations tolerate low gate error rates of about 0.1%). A quantum computer with such resources is referred to as *crypto-* *relevant quantum computer* (CRQC). At the time this paper has been written (2024), no such quantum computer is available. Note, that the quantum part computing elliptic curve discrete logarithms requires less quantum resources [RN+17].

Considering that the number of physical qubits needed to implement a single logical qubit is estimated to be about 1.000 [C22], and considering the corresponding gate errors, a quantum computer that can successfully perform the quantum part of Shor's algorithm seems to be far out. But recent advancements in quantum error correction indicate that orders of magnitudes fewer physical qubits are needed to realize a logical qubit (e.g. [BC+24], [ZS+23]). Thus, a crypto-relevant quantum computer may be sooner available than assumed a few years ago.

To be able to solve problems on today's error prone quantum computers (see [LB20]) so-called variational quantum algorithms are used [CA21]. These algorithms consist of a parameterized quantum circuit that produces a quantum state that is measured, this measurement result is evaluated by the classical part of the algorithms that optimizes the parameters, and the optimized parameter values are used by the parameterized quantum circuit again; this loop is performed until the overall procedure converges. Hereby, the quantum part runs only for a short period of time to avoid that the errors on the quantum computer pile up, and only few qubits are used - i.e. variational quantum algorithms are suitable for today's noisy and intermediate scale quantum (NISQ) computers [P18].

In [Y+23] the authors claim that their variational quantum algorithm can break RSA2024 with less than 400 qubits, although the number of operations of the quantum part of their algorithm is still to large. The authors of [KY23] implemented this algorithm but failed to scale it beyond 70 bit numbers where the algorithm always failed in their experiments. Thus, it is still open whether variational quantum algorithms can break security based on near-term quantum computers.

### 3.5. A Note on Symmetric Encryption

Shor's algorithm can break cryptographic schemes based on factorization or elliptic curves, which are asymmetric schemes. Symmetric schemes like AES 256 can be cracked by means of a quantum computer too via a brute-force attack [B05]: This is because Grover's quantum algorithm supports an unstructured search with a quadratic speedup [NC16].

But Grover's algorithm too requires an error corrected quantum computer. [W+22] presented a variational quantum algorithm to realize a brute-force attack on AES-like cryptographic schemes. In [WW+22] a (non-variational) quantum circuit is suggested that can break AES 128 with about 400 logical qubits.

Thus, symmetric cryptographic schemes are volatile to quantum attacks too. In contrast to asymmetric schemes, a symmetric scheme can be hardened against such a brute-force attack by doubling the key size used: since an attack based on the Grover algorithm has quadratic speedup, doubling the key size requires the same effort on a quantum computer that has to be spend on a classical computer for a key of the original key size. Under certain practical assumptions, less than doubling the key size suffice [F17].

# 4. Lattice-Based Cryptography

Cryptographic schemes that are based on lattices are less known than schemes based on factorization or elliptic curves. But they obey properties that make them candidates for quantum-resistant security mechanisms. In this chapter we sketch the basics of lattice-based cryptography. Basics about lattices can be found in [HNP21], and in addition, [P16] and [S20] discuss lattice problems and corresponding cryptography. Also, in [Z22] proofs for many of the facts given here can be found.

## 4.1. Basics About Lattices

A lattice is a set of points in a vector space that is build by taking all linear combinations with integer coefficients of a given set of linear independent vectors.

**Definition 10**: Let $v_1, \ldots, v_n \in \mathbb{R}^n$ be linear independent. The set

$$\Lambda < v_1, \ldots, v_n > := \left\{ \sum_{i=1}^{n} g_i v_i \mid g_i \in \mathbb{Z} \right\} \tag{17}$$

is called a *lattice* in $\mathbb{R}^n$ with basis $\{v_1, \ldots, v_n\}$. □

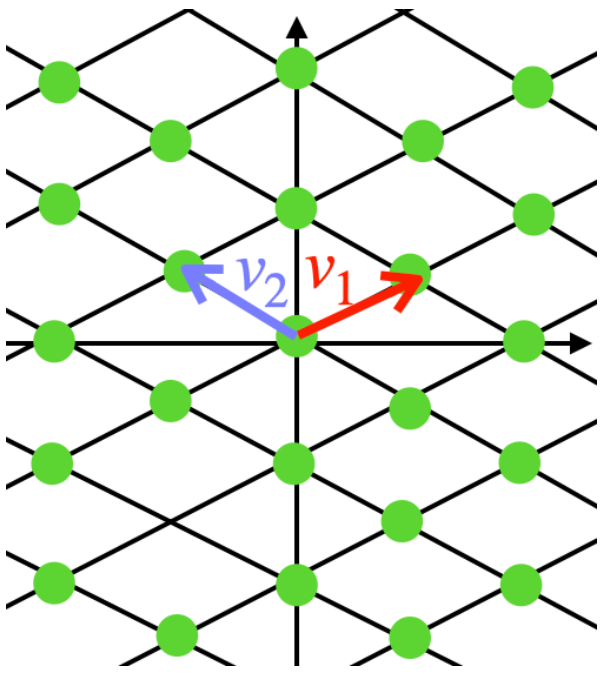

**Fig. 3**. A Sample Lattice

In the literature a generalized definition can be found with $k$ independent vectors $v_1, \ldots, v_k \in \mathbb{R}^n$; $k$ is then called the rank of the lattice, and for $k = n$ the lattice is called full rank. In what follows this generalization is not needed, thus, Definition 10 suffice. Figure 3 shows a lattice in $\mathbb{R}^2$.

It is obvious that a lattice is an abelian subgroup of $\mathbb{R}^n$. Furthermore, a lattice is discrete: for each lattice $\Lambda$ an $\varepsilon > 0$ can be found such that for all $x \neq y \in \Lambda$ it is $\| x - y \| \geq \varepsilon$, i.e. two points can be separated by disjoint neighborhoods. In fact, being a discrete abelian subgroup of $\mathbb{R}^n$ is equivalent of being a lattice in $\mathbb{R}^n$ [HNP21]:

**Note 6**: $\Lambda \subset \mathbb{R}^n$ is a lattice $\Leftrightarrow$ $\Lambda$ is a discrete abelian subgroup of $\mathbb{R}^n$. ∎

The following observation is important for lattice-based cryptography:

**Note 7**: Let $M \in \mathrm{GL}(n, \mathbb{R})$ (i.e. $M$ is an invertible $n \times n$ matrix). Then, the set $\{x \in \mathbb{Z}^n \mid Mx = 0\}$ is a lattice in $\mathbb{R}^n$.

Proof: Let $m_1, \ldots, m_n$ denote the columns of $M$, i.e. $M = (m_1 \ldots m_n)$. Because $M$ is invertible, $m_1, \ldots, m_n \subset \mathbb{R}^n$ are linear independent. Thus,

$$\{Mx \mid x \in \mathbb{Z}^n\} = \left\{ \sum x_i m_i \mid x_i \in \mathbb{Z} \right\} = \Lambda < m_1, \ldots, m_n >$$

is a lattice according to Definition 10.

It is obvious that $\{x \in \mathbb{Z}^n \mid Mx = 0\}$ is an abelian subgroup of $\mathbb{R}^n$. As a lattice, $\{Mx \mid x \in \mathbb{Z}^n\}$ is discrete. Obviously, a subset of a discrete set is discrete too, i.e.

$$\{x \in \mathbb{Z}^n \mid Mx = 0\} \subseteq \{Mx \mid x \in \mathbb{Z}^n\}$$

is discrete. Note 6 proves the claim. ∎

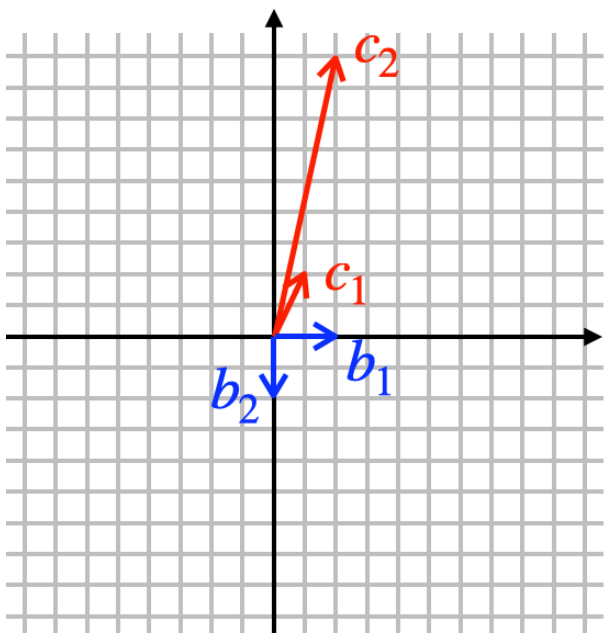

**Fig. 4**. Two Bases of a Lattice

## 4.2. Nearly Orthogonal Bases

A lattice has many different bases: let $\mathcal{B} = \{b_1, \ldots, b_n\}$ be a basis of the lattice $\Lambda$ and let B be the matrix with columns $b_1, \ldots, b_n$. For an $n \times n$ matrix $\mathcal{U}$ with integer coefficients (i.e. $\mathcal{U} \in \mathbb{Z}^{n \times n}$) and with $\det \mathcal{U} = \pm 1$ (a so-called *unimodular* matrix), the columns $c_1, \ldots, c_n$ of the matrix $C = B \cdot \mathcal{U}$ build a basis $\mathcal{C} = \{c_1, \ldots, c_n\}$ of $\Lambda$ (see the Figure 4 for an example).

The two bases shown in Figure 4 have quite different behavior in terms of computing solutions of lattice problems discussed below: a basis with very long vectors and a very small angle between vectors typically require higher computational effort than a basis with small vectors that are pairwise nearly orthogonal [S20].

The ideal basis of the lattice $\mathbb{Z}^n$ is an orthonormal basis $\mathcal{O} = \{o_1, \ldots, o_n\}$. It spans the unit cube the volume of which is given by the determinant $|\det O| = 1$ where $O = (o_1 \ldots o_n)$ is the matrix with columns $o_i$. Now, an arbitrary lattice $\Lambda$ is not generated by an orthonormal basis, but by a basis $\mathcal{B} = \{v_1, \ldots, v_n\}$. If $\{v_1, \ldots, v_n\}$ would be orthogonal (but not necessarily normalized), the volume of the cube spanned by $\{v_1, \ldots, v_n\}$ would simply be $\prod_{i=1}^n \| v_i \|$ . In case the vectors are not orthogonal they span a parallelepiped with volume $|\det B| = 1$ where $B = (v_1 \ldots v_n)$ is the matrix with columns $v_i$. Now, the "goodness" of the basis $\mathscr{B}$ can be measured by comparing the hypothetical cube spanned by orthogonal vectors of lengths $\| v_i \|$ and the volume $|\det B|$ of the parallelepiped spanned by $\mathscr{B}$:

**Definition 11**: Let $v_1, \ldots, v_n \in \mathbb{R}^n$ be linear independent, $\mathscr{B} = \{v_1, \ldots, v_n\}$ and $B = (v_1 \ldots v_n)$ the matrix with columns $v_i$. The number

$$\delta(\mathscr{B}) := \frac{\prod_{i=1}^{n} \| v_i \|}{|\det B|} \tag{18}$$

is called (*orthogonality*) *defect* of $\mathscr{B}$. □

For any matrix $M = (m_1 \ldots m_n)$ it is always $\det M \leq \prod_{i=1}^{n} \| m_i \|_2$ (Hadamard's inequality). Thus,

$$\frac{\prod_{i=1}^{n} \| m_i \|}{|\det M|} \geq 1$$

and consequently $\delta(\mathscr{B}) \geq 1$. If $\mathscr{B}$ is orthogonal then $\det B = \prod_{i=1}^{n} \| v_i \|$, i.e. $\delta(\mathscr{B}) = 1$ and vice versa: this explains the name "orthogonality" defect.

The *Lattice Reduction Problem* asks to determine a basis $\mathscr{B}$ for $\Lambda$ with minimal $\delta(\mathscr{B})$: This problem is NP hard. In practice, a *good basis* (not necessarily one with minimal orthogonality defect) of a lattice is chosen and becomes a private key, and a corresponding bad basis - which is computed based on a randomly chosen unimodular matrix - becomes the corresponding public key (see section 4.6).

Next, we present three basic lattice problems that are hard to solve and which are central to cryptography.

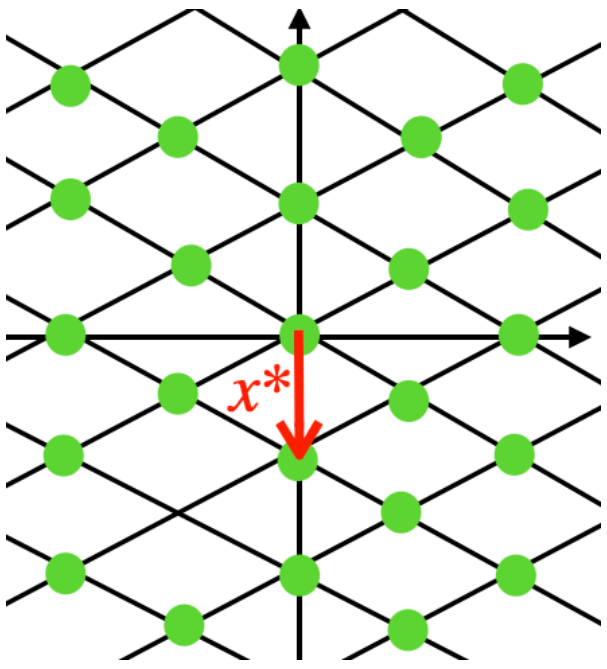

**Fig. 5**. The Shortest Vector of a Lattice

## 4.3. Shortest Vector

For a given lattice $\Lambda < v_1, \ldots, v_n >$ and a norm $\|\cdot\|$, the Shortest Vector Problem asks to find a non-zero vector of the lattice that has minimal length (see Figure 5).

**Definition 12**: Let $\Lambda < v_1, \ldots, v_n >$ be a lattice and $\| \cdot \|$ be a norm on $\mathbb{R}^n$. With $\lambda_1(\Lambda) := \min_{x \in \Lambda \setminus \{0\}} \| x \|$ the *Shortest Vector Problem* (SVP) asks to determine an $x^* \in \Lambda$ such that $\left\| x^* \right\| = \lambda_1(\Lambda)$. □

The importance of this problem is because of (see [M16])

**Theorem 4**: SVP is NP-hard for $\| \cdot \|_2$ and NP-complete for $\| \cdot \|_\infty$. ∎

### 4.4. Closest Vectors

For a given lattice $\Lambda < v_1, \ldots, v_n >$ , a norm $\|\cdot\|$ and a vector $w \in \mathbb{R}^n$, the Closest Vector Problem asks to find a vector of the lattice that is closest to $w$ (see Figure 6).

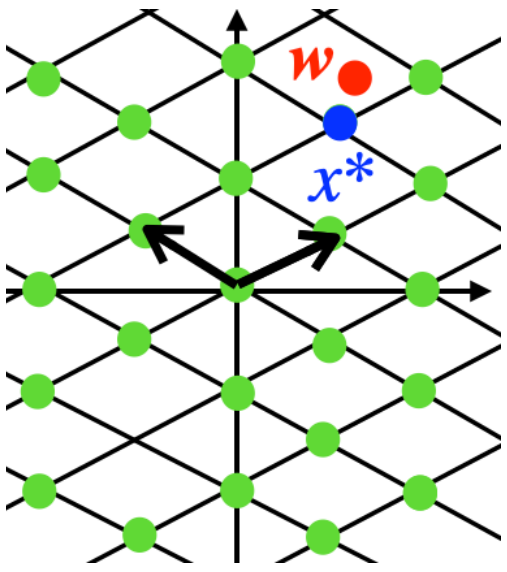

**Fig. 5**. The Vector of a Lattice Closest to a Given Vector

**Definition 13**: Let $\Lambda < v_1, \ldots, v_n >$ be a lattice, $\| \cdot \|$ be a norm on $\mathbb{R}^n$, and $w \in \mathbb{R}^n$. With $\operatorname{dist}(\Lambda, w) := \min_{x \in \Lambda} \| x - w \|$ the *Closest Vector Problem* (CVP) asks to determine an $x^* \in \Lambda$ such that $\left\| x^* - w \right\| = \operatorname{dist}(\Lambda, w)$. □

As before, the importance of this problem results from (see [M01], [M16])

**Theorem 5**: CVP is NP-complete. ∎

### 4.5. Shortest Independent Vectors

The Shortest Independent Vectors Problem is a bit more comprehensive: a basis of the lattice has to be found whose longest vector has minimal length across all bases of the lattice.

**Definition 14**: Let $\Lambda < v_1, \ldots, v_n >$ be a lattice and let $\mathfrak{B}$ be the set of all bases of $\Lambda$. With $M(\Lambda) := \min_{\{b_1, \ldots, b_n\} \in \mathfrak{B}} \max_{1 \leq i \leq n} \left\| b_i \right\|$ the *Shortest Independent Vectors Problem* (SIVP) asks to determine a basis $\{s_1, \ldots, s_n\}$ of $\Lambda$ with $\max_{1 \leq i \leq n} \| s_i \| = M(\Lambda)$. □

Again, the importance of this problem is because of (see [M16])

**Theorem 6**: SIVP is NP-complete. ∎

### 4.6. Example: Goldreich-Goldwasser-Halevi Cryptosystem

We sketch a public key cryptosystem that is based on the Closest Vector Problem (see [GGH97] and [G18] for more details). Let $\mathcal{B} = \{b_1, \ldots, b_n\}$ be a good basis of the lattice $\Lambda$ (i.e. $\delta(\mathcal{B}) \approx 1$), B be the matrix with columns $b_1, \ldots, b_n$, and let $\mathcal{U}$ be a unimodular matrix (see section 4.2). $\mathcal{B}$ is the private key of the cryptosystem. Next, the columns of $\hat{B} := \mathcal{U}B$ are a basis $\hat{\mathcal{B}}$ of the lattice $\Lambda$ with $\delta(\hat{\mathcal{B}}) \gg 1$ (i.e. $\hat{\mathcal{B}}$ is a bad basis). $\hat{\mathcal{B}}$ is the public key of the cryptosystem.

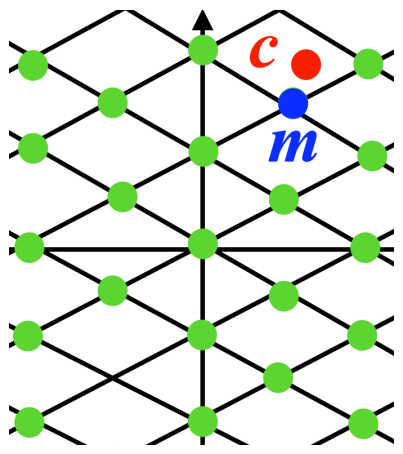

**Fig. 6**. Decrypting by Finding the Closest Vector to a Given Vector Representing a Massage

The sender encrypts a message $m$ as follows:

- The message $m$ is mapped to a lattice point
  - E.g. $m$ is split into $n$ blocks of characters and each block is mapped to an integer which is used as coefficient in a linear combination with the good basis) - see $m \in \Lambda$ in Figure 6.
- Randomly select a small vector $e \in \mathbb{R}^n$
- The encrypted message $c$ is $c = m\hat{B} + e$

The receiver decrypts $c$ by finding the vector closest to $c$ as follows:

- Compute $c \cdot B^{-1} = (m\hat{B} + e) \cdot B^{-1} = m\hat{B}B^{-1} + eB^{-1} = m\mathcal{U}\ BB^{-1} + eB^{-1} = m\mathcal{U} + eB^{-1}$
- Round the result according to [B86] to remove $eB^{-1}$, i.e. $m\mathcal{U}$ is attained
- Finally, $m = m\mathcal{U}\mathcal{U}^{-1}$

Note, that there is a well-known attack (Nguyen's attack) to this cryptosystem (see [G18]), thus, it should not be used in practice.

### 4.7. Average Case vs. Worst Case

Cryptography requires that randomly selected instances of problems are hard to solve [P16], which is referred to as *average-case hard*. In contrast, the theory of algorithms typically considers hardness in terms of the existence of instances of problems that are hard to solve, which is referred to as *worst-case hard*. For example, computing a prime factor of a number is worst-case hard but not average-case hard: randomly selecting a number will result in an even number with 0.5 probability and a prime factor "2" is immediately known.

The average-case hardness of certain lattice problems has been proven based on the worst-case hardness of some other lattice problems [A96]. As a consequence, cryptographic mechanisms can be designed that are provable average-cased hard unless all (!) instances of some other lattice problem can easily be solved. Learning with errors is such a provable average-case hard problem (see next section).

### 4.8. Learning With Errors

A *ring* $R$ is a set with two binary operations denoted by "+" and "·". In a ring, the pair $(R, +)$ is a group, the multiplication "·" is associative, and there is a multiplicative identity (or *unity*) "1". Both operations combined obey the laws of distributivity. The set of integers $\mathbb{Z}$ is the typical example of a ring. In a ring, elements typically have no multiplicative inverse, i.e. for an element $a \in R$ there is no element $b \in R$ such that $b \cdot a = 1$. If every element of the ring has a multiplicative inverse (and "·" is commutative) a ring is called a *field* (like the real numbers $\mathbb{R}$ or the complex numbers $\mathbb{C}$). See [L02] for the details.

Let $R$ be a finite ring, $a_1, \ldots, a_n \in R^n$ and $t_1, \ldots, t_n \in R$. Consider an unknown linear function $f : R^n \to R$ with $f(a_i) \approx t_i$, i.e. it is $f(a_i) = t_i$ but with small errors. The *Learning With Errors* (*LWE*) problem [R05] is to determine $f$ based on the set of training pairs $\{(a_1, t_1), \ldots, (a_n, t_n)\} \subset R^n \times R$.

This problem can be reformulated as follows: choose a fixed $s \in R^n$ ("the secrete"), and pick uniformly random small elements ("the errors") $e_1, \ldots, e_n \in R$. Then compute $t_i = \langle a_i, s \rangle + e_i$. Given $\{(a_1, t_1), \ldots, (a_n, t_n)\} \subset R^n \times R$, find $s$.

If $A = (a_1, \ldots, a_n) \in R^{n \times n}$ is defined as the matrix with columns $a_1, \ldots, a_n$, $t = (t_1, \ldots, t_n)^T \in R^n$ and $e = (e_1, \ldots, e_n)^T \in R^n$, then the LWE problem is seen as the problem to find a solution $s$ of the linear equation system

$$A \cdot s + e = t \qquad (19)$$

In case no errors occur (i.e. $e = 0$) Gaussian elimination solves the linear equation system $As = t$ in polynomial time. But when errors are to be considered, the LWE problem turns out to be surprisingly difficult: [R05] proves that if the LWE problem can be solved efficiently then solutions of the shortest vector problem SVP can be approximated by an efficient quantum algorithm. Since even the approximate SVP is assumed to be hard, the LWE problem is hard too (see [MR09], [R10]).

### 4.9. LWE as a Lattice Problem

In cryptographic applications the ring $R$ used in the LWE problem is $\mathbb{Z}_q$ for some prime number $q \in \mathbb{P}$: $\mathbb{Z}_q$ consists of the set of numbers $\{0,1,\ldots,q-1\}$. Two elements of $\mathbb{Z}_q$ are added and multiplied modulo $q$, i.e. for $a, b \in \mathbb{Z}_q$ it is $a + b := a + b \mod q$ and $a \cdot b := a \cdot b \mod q$. With these operations, $\mathbb{Z}_q$ is a finite commutative ring (see [MMS22] for details).

With $R$ substituted by $\mathbb{Z}_q$, $A \in \mathbb{Z}_q^{n \times n}$ and $s, t, e \in \mathbb{Z}_q^n$, the LWE problem in the shape of equation (19) can be written as

$$\left(A \,|\, E_n \,|\, -t\right) \cdot \begin{pmatrix} s \\ e \\ 1 \end{pmatrix} = 0 \qquad (20)$$

where $E_n \in \mathbb{Z}_q^{n \times n}$ is the identity matrix, and "|" denotes concatenation. According to Note 6, the set of integer solutions $\Lambda = \left\{ x \in \mathbb{Z}^{2n+1} \;\middle|\; \left(A \,|\, E_n \,|\, -t\right) \cdot x = 0 \right\}$ is a lattice. Obviously, $(s, e, 1)^T \in \mathbb{Z}^{2n+1}$ is such a solution, i.e. a lattice point. Especially, it can be shown that $(s, e, 1)^T \in \Lambda$ solves the shortest vector problem SVP for $\Lambda$, i.e. solving the LWE problem means to solve the SVP:

**Note 8**: The Learning With Errors problem is a lattice problem. ∎

Thus, cryptographic schemes based on the LWE problem belong to the domain of lattice-based cryptography. [MR09] provides examples of lattice-based schemes for encryption and signature. We sketch a corresponding encryption schema in section 5.4.

## 5. Algebraic Lattice Cryptography

Key sizes in LWE get large (see section 4.6 for an example), they grow in the order of $n^2$ [R10]. Large keys are not acceptable in several application domains. Reducing these key sizes to grow linearly in $n$ can be achieved by "adding more algebraic structure" to the lattice. More concrete, the group $\mathbb{Z}_q$ is substituted by the ring of polynomials $\mathbb{Z}_q[X]/(X^n + 1)$ without loosing the hardness property of regular LWE [R10]. Going even further, the ring of polynomials $\mathbb{Z}_q[X]/(X^n + 1)$ is extended to the module $(\mathbb{Z}_q[X]/(X^n + 1))^k$.

In this chapter, we provide the basics about polynomials and $\mathbb{Z}_q[X]/(X^n + 1)$, and also about modules (see [L02] for more algebraic depth). We sketch the use of this algebraically extended lattices (also called *algebraic lattices*) in cryptography ([MR09] and [R10] provide further details). Based on such lattices Kyber as an encryption mechanism and Dilithium as a signature mechanism are sketched.

### 5.1. Polynomial Rings

When substituting $\mathbb{Z}_q$ by the ring $\mathbb{Z}_q[X]/(X^n + 1)$ the resulting LWE problem is called *ring-LWE* problem [LP+10]. Beyond that, when the ring $\mathbb{Z}_q[X]/(X^n + 1)$ is even substituted by the module $(\mathbb{Z}_q[X]/(X^n + 1))^k$. The corresponding LWE problem is called *module-LWE*.

The advantage by moving from ring-LWE to module-LWE is as follows [MA+23]: if the security level has to be increased in ring-LWE, the degree $n$ of the polynomials must be increased which in turn increases the computational effort involved. In module-LWE the degree of the polynomials can remain the same but the rank $k$ of the module has to be enlarged. The advantage is that any optimization in manipulating polynomials remains the same.

Thus, we must understand the basics of polynomials and how to perform computations with them.

**Definition 15**: Let $R$ be a commutative ring. A *polynomial* in $X$ over $R$ is an expression $p = a_0 + a_1X + \ldots + a_nX^n$ with $a_i \in R$; each $a_i$ is called a *coefficient* of the polynomial. With $a_n \neq 0$ the number $n$ is called the *degree* of the polynomial: $n = \deg p$. $R[X]$ denotes the set of all polynomials in X over $R$. □

Polynomials can be added and multiplied: For $p = \sum_{i=0}^{n} a_iX^i$ and $q = \sum_{i=0}^{m} b_iX^i$ in $R[X]$ we define the sum $p + q$ of the two polynomials as

$$p + q = \sum_{i=0}^{\max(m,n)} r_iX^i \text{ with } r_i = a_i + b_i \tag{21}$$

and the product $p \cdot q$ of the two polynomials as

$$p \cdot q = \sum_{i=0}^{m+n} s_iX^i \text{ with } s_i = \sum_{j+k=i} a_jb_k \tag{22}$$

With these two operations on $R[X]$ the following can be proven:

**Lemma 4**: $(R[X], +, \cdot)$ is a commutative ring. ∎

In applications of algebraic lattices like Kyber (see section below) messages are considered as bit strings and they are transformed into polynomials as follows: for a message $m = b_0 \ldots b_n$ with $b_i \in \{0,1\}$ and $b_n \neq 0$, the polynomial from $\mathbb{Z}[X]$ representing $m$ is $P(m) = b_0 + b_1X + \cdots + b_nx^n$. When adding and multiplying polynomials their coefficients become large, especially when their degree (which corresponds to the number of bits of a message) is large - see formula (22). To avoid inefficiency in the corresponding computations with large numbers, all computations are taken $\bmod q$ for a prime number $q \in \mathbb{P}$. Thus, the set of polynomials $\mathbb{Z}_q[X]$ is used which ensures that all coefficients of polynomials are less then $q$.

## 5.2. Residue Classes of Polynomials

Also, when multiplying polynomials the degree of the resulting polynomial gets large (i.e. the degrees are summed up - see equation (22)): this results in inefficiencies too. Consequently, the degree of polynomials is limited by using residue classes of polynomials. If $f$ is an *irreducible* polynomial (i.e. $f$ is not the product of two non-constant polynomials) then the set of all residue classes $\mathbb{Z}_q[X]/f$ is again a commutative ring. The polynomial $1 \cdot X^0$ is the identity element in $\mathbb{Z}_q[X]/f$. Furthermore, if $\deg f = n$ then $\deg g \leq n - 1$ for each $g \in \mathbb{Z}_q[X]/f$. Thus, $\mathbb{Z}_q[X]/f$ contains $n \cdot q$ elements, i.e. $\mathbb{Z}_q[X]/f$ is a finite ring. All hardness properties of the regular LWE problem are inherited by the ring-LWE problem over the ring $\mathbb{Z}_q[X]/f$. In summary:

**Lemma 5**: Let $f \in \mathbb{Z}_q[X]$ be irreducible. Then, $\mathbb{Z}_q[X]/f$ is a finite commutative ring with unity and card $\mathbb{Z}_q[X]/f = q \cdot \deg f$. ∎

Computing the residue class of a polynomial $g \in \mathbb{Z}_q[X]$ requires to divide $g$ by $f$ and take the remainder of the division as the result. We briefly remind polynomial division by means of the following example:

- Let $f = x^2 + 1$ be an irreducible polynomial in $\mathbb{Z}[X]$
- Let $p = 4x^5 - x^4 + 2x^3 + x^2 - 1 \in \mathbb{Z}[X]$
- By computing $(4x^5 - x^4 + 2x^3 + x^2 - 1) \div (x^2 + 1)$ and determining the remainder of this division, $p \mod f \in \mathbb{Z}[X]/(X^2 + 1)$, is computed
- The computation is in analogy to the division of numbers (see Figure 7)

$$\left(4x^5 - x^4 + 2x^3 + x^2 - 1\right) \div \left(x^2 + 1\right) = 4x^3 - x^2 - 2x + 2 + \frac{2x - 3}{x^2 + 1}$$

$$\begin{array}{l}
\underline{-4x^5 \quad -4x^3} \\
\quad -x^4 - 2x^3 \\
\quad \underline{+x^4 \quad +x^2} \\
\qquad -2x^3 + 2x^2 \\
\qquad \underline{+2x^3 \quad +2x} \\
\qquad\quad 2x^2 + 2x \\
\qquad\quad \underline{-2x^2 \quad -2} \\
\qquad\qquad 2x - 3
\end{array}$$

**Fig. 7**. Example of a Polynomial Division

- Determine the factor of $x^2$ of the denominator to result in the highest term $4x^5$ of the numerator of the division: this is $4x^3$, the first summand of the result of the division
- Multiply $4x^3$ by the denominator $x^2 + 1$ which results in $4x^5 + 4x^3$
- Subtract $4x^5 + 4x^3$ this from the numerator
- The highest term of the remaining polynomial is $-x^4$
- Determine the factor of $x^2$ of the denominator to result in this highest term $-x^4$ of the remaining numerator: this is $-x^2$, the second summand of the result of the division
- Multiply $x^2$ by $x^2 + 1$ which results in $x^4 + x^2$ …
- …
- The remainder of this polynomial division is $2x - 3$
- Thus, $(4x^5 - x^4 + 2x^3 + x^2 - 1) \equiv (2x - 3) \in \mathbb{Z}[X]/(X^2 + 1)$

Note, that the coefficients of the polynomials in the example are from $\mathbb{Z}$. Computations in $\mathbb{Z}_q[X]/(X^2 + 1)$ must perform all computations involving coefficients of the polynomials $\mod q$.

### 5.3. Modules of Polynomials

As mentioned above, increasing the security level of a cryptographic mechanism based on ring-LWE over $\mathbb{Z}_q[X]/f$ requires to increase the degree $n$ of the polynomials, which in turn increases the involved computational effort. To avoid this, the ring $\mathbb{Z}_q[X]/f$ is substituted by $(\mathbb{Z}_q[X]/f)^k$; the latter algebraic structure is called a module.

**Definition 16**: A *module* $M$ over a ring $R$ is a commutative group $(M, +)$ with an operation $\cdot : R \times M \to M$ of $R$ on $M$ such that the laws of distributivity are satisfied, i.e. $(r_1 + r_2) \cdot m = r_1 m + r_2 m$ and $r \cdot (m_1 + m_2) = r \cdot m_1 + r \cdot m_2$. □

If instead of a ring $R$ a field $\mathbb{K}$ is used in the definition, the module over $\mathbb{K}$ becomes a vector space over $\mathbb{K}$. Thus, modules are generalizations of vector spaces. But in general, modules are quite different from vector spaces, e.g. not every module has a basis, and if a module has a basis then different of its bases may have different cardinalities.

The cartesian product $R^k$ of a ring $R$ is a module over the ring $R$: elements of $R^k$ are added component-wise, and multiplication of an element of $R^k$ with an element of $R$ is done component-wise too - just like vectors in the coordinate space $\mathbb{K}^k$ are added and multiplied with scalars from $\mathbb{K}$. Thus, based on Lemma 5 we get

**Lemma 6**: Let $f \in \mathbb{Z}_q[X]$ be irreducible, $k \in \mathbb{N}$. Then, $(\mathbb{Z}_q[X]/f)^k$ is a finite module with card $(\mathbb{Z}_q[X]/f)^k = (q \cdot \deg f)^k$. ∎

$(\mathbb{Z}_q[X]/f)^k$ is a finite module because card $\mathbb{Z}_q[X]/f = q \cdot \deg f$ according to Lemma 5, thus card $(\mathbb{Z}_q[X]/f)^k = (q \cdot \deg f)^k < \infty$.

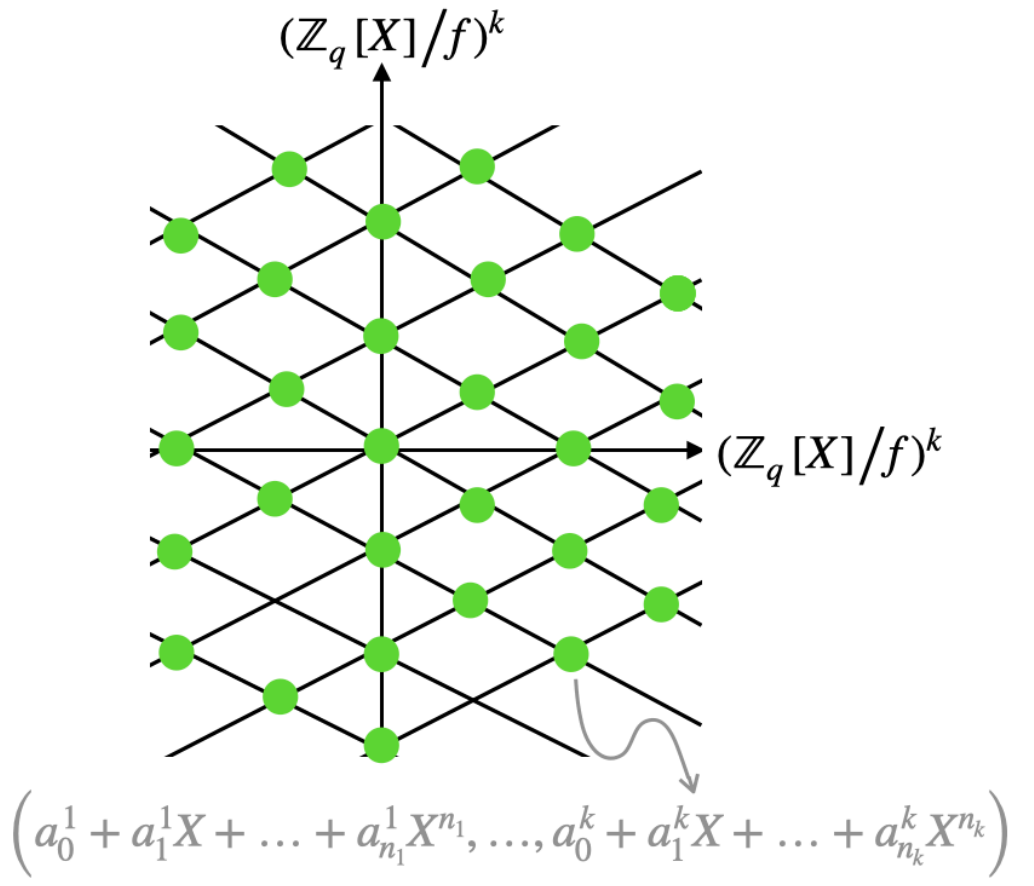

**Fig. 8**. Geometric Situation of Algebraic Lattices

The notion of a lattice can be generalized to the module $(\mathbb{Z}_q[X]/f)^k$. For this purpose, $\mathbb{Z}_q[X]/f$ is considered as a set of tuples by mapping a polynomial $a_0 + a_1 X + \ldots + a_m X^m$ to the tuple $(a_0, a_1, \ldots, a_m, 0, \ldots, 0) \in \mathbb{Z}_q^{\deg f}$, i.e. the tuple consisting of the coefficients of the polynomial. This mapping is performed for each component polynomial $p_i$ of $(p_1, \ldots, p_k) \in (\mathbb{Z}_q[X]/f)^k$ and the tuples are combined. This way, each element of $(\mathbb{Z}_q[X]/f)^k$ becomes an element of $\mathbb{Z}_q^{k \cdot \deg f} \subset \mathbb{Z}^{k \cdot \deg f}$ where lattices (of rank $k \cdot \deg f$) can be defined.

In order to be able to formulate the lattice problems introduced in sections 4.3, 4.4, and 4.5 a norm is needed on $(\mathbb{Z}_q[X]/f)^k$. But the canonical scalar product $\langle (a_i), (b_i) \rangle = \sum a_i b_i$ induces the canonical norm $\|(a_i)\| = \sqrt{\langle (a_i), (a_i) \rangle}$. Thus, lattice problems in $(\mathbb{Z}_q[X]/f)^k$ result.

Figure 8 depicts this situation: The module which is the foundation of an algebraic lattice is a cartesian product of a finite polynomial ring. Accordingly, the elements of the module are tuples of polynomials. Computations in this module are performed mod $q$ w.r.t. the coefficients of the polynomials, and mod $f$ w.r.t. the polynomials themselves. Different algebraic lattice-based cryptographic mechanisms (like Kyber or Dilithium) use different prime numbers $q$, different irreducible polynomials $f$, and different ranks $k$.

## 5.4. Kyber

Kyber is an algebraic-lattice-based asymmetric cryptographic mechanism based on a module-LWE problem described above. The specification of Kyber can be found in [AB+21]: it specifies all algorithms used for performing the calculations with polynomials, for rounding, for sampling error vectors etc. Kyber is using $R_q := \mathbb{Z}_q[X]/(X^n + 1)$ with $n = 256$ and $q = 3329$.

### 5.4.1. Computing Keys

The private and public key are determined as follows:

1. A randomly chosen $s \in R_q^k = (\mathbb{Z}_q[X]/(X^n + 1))^k$ becomes the *private key*. Here, Kyber is using $k \in \{2,3,4\}$ as the *key length*. This key length varies the lattice rank $k \cdot n$ and, thus, the security level implied. For example, with $k = 3$ the lattice rank becomes 768 and the corresponding variant KYBER768 has the security level of AES192.
2. Next, a matrix $A \in R_q^{k \times k}$ is randomly chosen. The module-LWE requires a small randomly generated error vector $e \in R_q^k$, and $t = As + e \in R_q^k$ is computed. The pair $(A, t)$ becomes the *public key*.

### 5.4.2. Encryption

For each message $m$ to be encrypted the following is performed:

1. First, the following elements are randomly generated: a so-called randomizer vector $r \in R_q^k$ as well as an error vector $e_1 \in R_q^k$ and a single error polynomial $e_2 \in R_q$, both with small coefficients.

2. The message $m$ is considered as a bit string $m = b_0 \ldots b_{n-1}$ and is represented as a polynomial $P(m) = b_0 + b_1 X + \cdots + b_{n-1} X^{n-1}$ (i.e. $b_i \in \{0,1\}$); in this sense, $m$ and $P(m)$ are used interchangeably.

3. $P(m)$ is "scaled" by multiplying it by $\lfloor q/2 \rceil$, i.e. $\lfloor q/2 \rceil \cdot P(m)$ is computed. As a reminder, $\lfloor x \rceil$ is the integer closest to $x$, e.g. $\lfloor 2.4 \rceil = 2$ or $\lfloor 2.5 \rceil = 3$. Note, that the coefficients of the latter polynomial are 0 or $\lfloor q/2 \rceil$ because the coefficients of $P(m)$ are 0 or 1.

4. Then, $u = A^T r + e_1$ and $v = t^T r + e_2 + m$ are computed.

5. The encrypted message is $(u, v)$.

5.4.3. Decryption

The encrypted message is $(u, v)$ is decrypted as follows:

1. First, $\widehat{m} = v - s^T u$ is computed by using the secret key $s$.
   - Note, that $\widehat{m}$ is not yet $m$, but it contains some noise:

$$\begin{aligned}
\widehat{m} = v - s^T u &= \left(t^T r + e_2 + m\right) - s^T\left(A^T r + e_1\right) \\
&= \left((As + e)^T r + e_2 + m\right) - s^T\left(A^T r + e_1\right) \\
&= s^T A^T r + e^T r + e_2 + m - s^T A^T r - s^T e_1 \\
&= e^T r + e_2 + m - s^T e_1 \\
&= m + \underbrace{e^T r + e_2 - s^T e_1}_{\text{noise}}
\end{aligned} \tag{23}$$

2. Since $e, e_1, e_2$ are "small" by definition, the noise $e^T r + e_2 - s^T e_1$ has "small" coefficients too. Thus, the coefficients of $\widehat{m}$ are close to the coefficients of $\lfloor q/2 \rceil \cdot P(m)$ which are 0 or $\lfloor q/2 \rceil$ (see 5.4.2. (3)).

3. Next, the following "rounding" of the coefficients of $\widehat{m}$ is performed resulting in $\rho(\widehat{m})$:
   - All coefficients of $\widehat{m}$ close to $\lfloor q/2 \rceil$ are set to "1".
   - All coefficients of $\widehat{m}$ close to 0 (or close to q, which is congruent 0) are set to "0".

4. Finally, $\rho(\widehat{m}) = m$ is (with very high probability) the original message.

There is a small probability that $\rho(\widehat{m})$ is not the original message $m$, but this failure probability is extremely small (see [BD+18], Theorem 1): for example, for $k \cdot n = 512$ the failure probability is $2^{-139}$.

[BD+18] also explains the reasons for the hyper-parameters chosen for Kyber in more detail. The rounding mechanism is in fact more sophisticated than presented here and is described in [B86].

5.4.4. Kyber by Example

In what follows, we describe Kyber stepwise by example (remark: no guarantee that the computations [involving lots of polynomial multiplications, frequently taking coefficients mod 7 and polynomials mod $(X^4+1)$] are error free!). We use small hyperparameters to highlight the essential processing: As hyperparameters we set $n=4$, $q=7$ and $k=2$. Thus, our finite polynomial ring is $R_7=\mathbb{Z}_7[X]/(X^4+1)$.

First, we need to choose a private key and compute a corresponding public key:

- As private key we choose $s=\begin{pmatrix} x^3+x+1 \\ x+2 \end{pmatrix}$
- $A\in R_7^{2\times 2}$ is chosen randomly as $A=\begin{pmatrix} 4x^3+5x+4 & 3x^3+5x^2 \\ 5x^3+3x & 6x^2+6 \end{pmatrix}$
- In order to computed $t=As+e$ we need a small error vector $e\in R_7^2$ and choose $e=(x^2,x)$
- The computation results in (considering the coefficients mod 7 and the polynomials mod $(X^4+1)$)

$$t=\begin{pmatrix} 4x^3+5x+4 & 3x^3+5x^2 \\ 5x^3+3x & 6x^2+6 \end{pmatrix}\cdot\begin{pmatrix} x^3+x+1 \\ x+2 \end{pmatrix}+\begin{pmatrix} x^2 \\ x \end{pmatrix}=\begin{pmatrix} 5x^3-2x^2+2x-1 \\ 4x^3-4x^2+3x+4 \end{pmatrix}$$

- Thus, the public key $(A,t)$ is

$$\left(\begin{pmatrix} 4x^3+5x+4 & 3x^3+5x^2 \\ 5x^3+3x & 6x^2+6 \end{pmatrix},\begin{pmatrix} 5x^3-2x^2+2x-1 \\ 4x^3-4x^2+3x+4 \end{pmatrix}\right)$$

The message to encrypt is $\mu$ = ITS ALL GREEK TO ME (see section 2.4). For simplicity, we only consider the first letter, i.e. $m=9$ and we get $m=1001$ in binary representation. Next, $m$ must be mapped to a polynomial by means of the map $P$ resulting in

$$P(m)=x^3+0x^2+0x+1=x^3+1;$$

scaling this polynomial by $\lfloor q/2\rceil=\lfloor 7/2\rceil=4$ gives $m=4x^3+4$ (as before, we identify $m$ and $P(m)$).

- To encrypt $m$ we choose $r=(x^2,1)^T\in R_7^2$ as randomizer, an error vector $e_1=(x+1,1)^T\in R_7^2$ and as error polynomial $e_2=x^3+x\in R_7$.
- Then we compute $u=A^Tr+e_1$considering mod 7 w.r.t. the coefficients and mod $(X^4+1)$ w.r.t. the polynomials:

$$u=\begin{pmatrix} 4x^3+5x+4 & 5x^3+3x \\ 3x^3+5x^2 & 6x^2+6 \end{pmatrix}\begin{pmatrix} x^2 \\ 1 \end{pmatrix}+\begin{pmatrix} x+1 \\ 1 \end{pmatrix}=\begin{pmatrix} 3x^3+4x^2+1 \\ 6x^2-3x-5 \end{pmatrix}$$

- Next, we compute $v=t^Tr+e_2+m$ (again based on the residue classes for the coefficients and the polynomials):

$$v=\left(5x^3-2x^2+2x-1\ ,\ 4x^3-4x^2+3x+4\right)\begin{pmatrix} x^2 \\ 1 \end{pmatrix}+(x^3+x)+(4x^3+4)$$

$$=3x^3+2x^2-x-2$$

- Thus, the encrypted message is:

$$(u, v) = \left( \begin{pmatrix} 3x^3 + 4x^2 + 1 \\ 6x^2 - 3x - 5 \end{pmatrix}, 3x^3 + 2x^2 - x - 2 \right)$$

In order to decrypt the message, the recipient must first compute $\widehat{m} = v - s^T u$ based on the private key $s$. Thus:

$$\widehat{m} = (3x^3 + 2x^2 - x - 2) - \left(x^3 + x + 1,\ x + 2\right) \begin{pmatrix} 3x^3 + 4x^2 + 1 \\ 6x^2 - 3x - 5 \end{pmatrix} = 3x^3 - x^2 - x - 4$$

Computing the coefficients mod 7 results in

$$\widehat{m} = 3x^3 + 6x^2 + 6x + 3$$

and finally rounding the coefficients by $\rho$, i.e. coefficients close to $\lfloor q/2 \rceil = 4$ are set to "1", and coefficients close to 0 (or close to $q = 7$) are set to "0" delivers

$$\rho(\widehat{m}) = x^3 + 1 = 1 \cdot x^3 + 0 \cdot x^2 + 0 \cdot x + 1$$

Thus, $\rho(\widehat{m}) = P(m)$ and the original message $m = 1001$ results.

#### 5.4.5. Kyber as KEM

Like ECC Diffie-Hellman discussed in section 2.6, Kyber is used as a key encapsulation mechanism (KEM): The asymmetric encryption/decryption protocol described before is mainly used to agree on a symmetric key. Once the symmetric key is agreed on, the information exchanged is encrypted and decrypted based on the shared secrete.

### 5.5. Dilithium

We sketch the essence of Dilithium in this section. Note, that the details are more subtle than sketched (e.g. "rounding" means to use high-order bits of the coefficients of the affected polynomials); the details can be found in [BD+21]. [RB+22] summarizes Dilithium as well as other post-quantum cryptography algorithms.

Like Kyber, Dilithium is an algebraic-lattice-based cryptographic mechanism. But instead of encryption/decryption it is about signatures. Dilithium is also using $R_q = \mathbb{Z}_q[X] / (X^n + 1)$ with $n = 256$ but $q$ is much bigger in Dilithium than in Kyber: $q = 2^{23} - 2^{13} + 1 = 8380417$.

#### 5.5.1. Key Generation

Key generation works like in Kyber: a uniformly random $s \in R_q^k$ with small coefficients becomes the *private key*. Next, a uniformly random $A \in R_q^{m \times k}$ as well as a uniformly random $e \in R_q^m$ with small coefficients is chosen; based on the latter two $t = As + e \in R_q^m$ is computed. The pair $(A, t)$ is the *public key*. $m$ and $k$ are chosen in Dilithium in order to result in different security levels [JM+24], i.e. $(m, k) = (4,4)$ for security level 2, $(m, k) = (6,5)$ for security level 3, and $(m, k) = (8,7)$ for security level 5.

5.5.2. Signature Generation

Signatures are computed as follows:

1. For each message $\mu$ to sign, freshly choose uniformly random $r_1 \in R_q^k$ with small coefficients.
2. Compute $w = \rho\left(A r_1\right) \in R_q^m$
   - $\rho$ is the algorithm rounding coefficients to 0 or $\lfloor q/2 \rceil$
3. Compute $c = \Psi\left(\mu \| w\right) \in R_q$
   - $\Psi$ is a hash function based on the functions SHAKE 256 (which is variant of the SHA-functions [N15]) as well as SHA-256 itself [DK+18].
   - The hash function maps its argument to $B_h \subset \mathbb{Z}_q[X]/(X^n + 1)$ which is a set of polynomials with coefficients in $\{-1, 0, +1\}$ and exactly $h$ coefficients not being "0". Dilithium is using $h = 60$ resulting in more than $2^{256}$ potential polynomials.
4. Compute $r_2 = r_1 + cs \in R_q^k$
5. The *signature* is $(r_2, c)$. It is send together together with the message $\mu$ to the recipient. The recipient can verify the signature as discussed in section 5.5.3.

[B19] proves that Dilithium is a so-called self-target Module Short Integer Solution Problem (MSISP), i.e. a lattice problem: Given a message $\mu$, find $y \in R_q^{2k+1}$ with $0 < \| y \| \le \beta$ where $y = (r_1, c, r_2)^T$ with $r_1 \in R_q^k$ , $c \in B_{60}$ , $r_2 \in R_q^k$ such that $\Psi\left(\mu \, \| \, \left(A \,|\, s \,|\, E_k\right) \cdot y\right) = c \in B_{60}$. Here, $\left(A \,|\, s \,|\, E_k\right)$ is the matrix consisting of the matrix $A$ (part of the public key) followed by a single column $s$ (the private key) followed by the identity matrix $E_k$; the similarity to equation (20) already roughly indicates that this is a lattice problem.

5.5.3. Signature Verification

When the signed message $\left(\mu \left(r_2, c\right)\right)$ is received, the following computations are performed to verify the signature (the correctness of this verification procedure is shown in the next section):

1. Compute $\hat{w} = \rho\left(A r_2 - tc\right)$.
   - $(A, t)$ is the public key.
   - $(r_2, c)$ is the signature to verify.
2. Compute $\hat{c} = \Psi\left(\mu \| \hat{w}\right)$.
   - $\mu$ is the message claimed to be signed.
3. In case $\hat{c} = c$ then the signature is valid, otherwise the signature is invalid.

5.5.4. Correctness of Signature Verification

The following shows that the signature verification procedure is correct:

$$\begin{aligned}\rho\left(Ar_2 - tc\right) &\overset{(a)}{=} \rho\left(A(r_1 + cs) - (As + e)c\right)\\ &\overset{(b)}{=} \rho\left(Ar_1 + Acs - Asc - ec\right)\\ &\overset{(c)}{=} \rho\left(Ar_1 - ec\right)\\ &\overset{(d)}{=} \rho\left(Ar_1\right)\\ &\overset{(e)}{=} w\end{aligned}$$

The first equality (a) results from substituting $r_2 = r_1 + cs$ and $t = As + e$. (b) is a simple multiplication, followed by cancelling the two terms in the middle in (c). (d) is valid because $e$ has been chosen with small coefficients while $c \in B_{60}$ has coefficients from $\{-1, 0, +1\}$, i.e. they are small too, thus, $ec$ has small coefficients which implies that rounding $\rho$ is not affected by $ec$. (e) uses the definition of $w$. The verification succeeds.

Assume that $r_2$ has not been calculated with the correct private key $s$ (i.e. the substitution in (a)), i.e. the signature is bogus. But $t$ as part of the public key has been computed with the correct private key $s$. Consequently, the two terms in the middle of (b) would not cancel out and the equation in (c) would not hold. Thus, $\rho\left(Ar_2 - tc\right)$ would result in a value $\hat{w} \neq w$. This implies that the step (2) in the verification process results in $\hat{c} = \Psi\left(\mu \| \hat{w}\right) \neq c$: the verification fails.

## 6. Why Should I Care?

There is whole spectrum of opinions or estimations, respectively, about when powerful quantum computers will be available (and some even argue that such computers can not be build at all). The first question to ask is: “powerful for which purpose”. It is important to note that different application areas require different characteristics of a quantum computer (in terms of number of qubits, decoherence time, gate fidelity etc) to qualify as “powerful”. In our context, a quantum computer that can threaten today’s cryptographic schemes by being able to execute Shor’s algorithm, for example, is referred to as *Cryptographic Relevant Quantum Computer* (CRQC) (see section 3.4).

[SW+24] estimates that quantum computers powerful enough for several application areas will be available “in the late 2020” and that cryptographic relevant quantum computers are (a bit) further out. However, CRQC are threaten security already now (see section 6.1). The time a particular company has the counter this threat can be roughly estimated (see section 6.2). Awareness of this threat is growing in society (section 6.3), and standards for a quantum-safe infrastructure are being created (sections 6.4 and 6.5).

### 6.1. “Harvest Now, Decrypt Later”

Cryptographic relevant quantum computers are not only threaten security once they are available, but they do already now. An often perceived attack is referred to as “Harvest Now, Decrypt Later” (HNDL): here, an attacker can steal data that is encrypted based on an asymmetric scheme, store it, and decrypt it once a CRQC is available. If data that must be kept confidential and that is persistently stored is encrypted symmetrically, such data at rest is considered quantum-safe (in the sense of section 3.5). However, data at rest that is asymmetrically encrypted is in danger.

Data in motion (e.g. messages or documents exchanged) is typically encrypted based on a symmetric session key which has been agreed upon at start of the session based on asymmetric encryption. If this message exchange is captured and stored, the data in motion can be decrypted at a later point in time once a CRQC is available: the asymmetric keys will be decrypted allowing to derive the symmetric key to decrypt the proper data.

### 6.2. Mosca’s Inequality

Thus, companies should take proper actions to protect themselves from attacks by a CRQC. These actions include migrating to a cryptographic infrastructure that is quantum-safe. In Figure 9, this time is called $T_{Migrate}$. During this time period, and especially at the end of this period, data is very likely generated by the company that must be kept confidential for a company-specific time, called $T_{Confi}$ in Figure 9. This means that a company must ensure that encrypted data can not be decrypted for a period of $T_{Migrate} + T_{Confi}$.

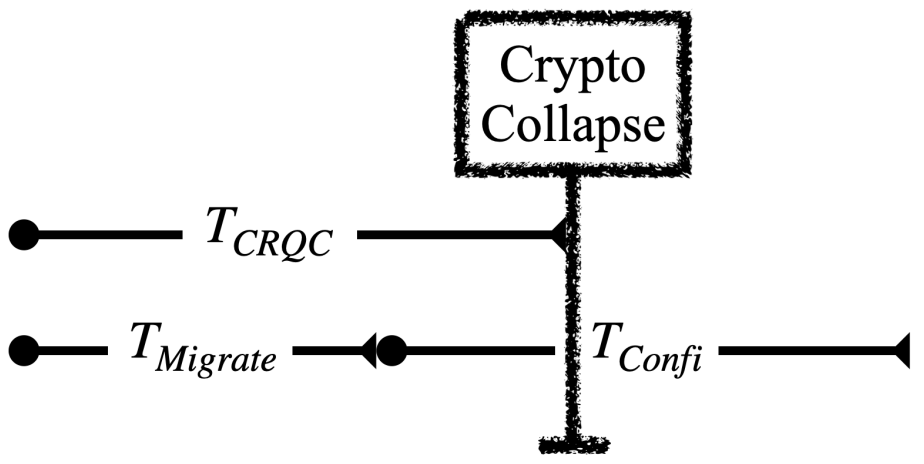

**Fig. 9**. Graphics of Mosca’s Inequality

The time needed to develop a CRQC is denoted by $T_{CRQC}$ in the figure. Once $T_{CRQC}$ is passed, data is at risk and can be decrypted: a “crypto collapse” may occur. While $T_{Confi}$ is typically well-known by each company itself, and $T_{Migrate}$ can be estimated by project management techniques of a company, $T_{CRQC}$ is out of control of a company. Roadmaps of vendors of quantum computers have be observed and tracked to derive $T_{CRQC}$. Also, assessments of consulting firms can be considered.

Once the numbers are available, the following inequality known as *Mosca's Inequality* [MP23] can be evaluated:

$$T_{Migrate} + T_{Confi} > T_{CRQC} \tag{24}$$

In case the inequality (24) holds, a company may get into trouble: the time data must be kept secrete and the time needed to move towards a post-quantum cryptographic infrastructure is bigger than the time needed for (some) vendors to provide a cryptographic relevant quantum computer. Thus, the data of the corresponding company is at risk. If possible, the migration time $T_{Migrate}$ may be shortened in order to make $T_{Migrate} + T_{Confi} \leq T_{CRQC}$ hold. But in case already $T_{Confi} > T_{CRQC}$ holds, the company is in trouble.

### 6.3. Awareness

Considerations like the ones outlined in the section before also apply to governments and policy makers in general. As a result the US government, for example, enacted a law [U22] "…to encourage the migration of Federal Government information technology systems to quantum-resistant cryptography…". In reaction to this law a strategy to migration to post-quantum cryptography has been worked out [U24], and the cost to migrate the IT systems of the US government has been estimated to be higher than 7 billion US$ by 2035.

Similarly, the European Commission recommends a coordinated implementation roadmap for the transition to post-quantum cryptography for its member states [EC24] "…necessary for Europe to look for stronger safeguards, ensuring the protection of sensitive communications and the long-term integrity of confidential information, i.e., by switching to Post-Quantum Cryptography as swiftly as possible". Furthermore, the intend for this roadmap is "…to foster the transition to Post-Quantum Cryptography for the protection of digital infrastructures and services…".

[DP23] is a study from 2023 about how industrial organizations worldwide are addressing post-quantum-safeness. About 1500 practitioners from the US, EU, and AP have been surveyed who claimed to be acquainted with their organizations work in the area of post-quantum cryptography. The result is that 61% of the practitioners are very much concerned about their organizations not being prepared to address the security threats posed by quantum computing. Even 74% are concerned about "harvest now, decrypt later" attacks. 23% of the respondents have a strategy towards post-quantum-safeness, and the same percentage (23%) have even no plan to build such a strategy.

In summary, it seems to be that government organizations take post-quantum security more serious than industrial organizations.

### 6.4. Standardization

Standards are important for post-quantum security. For example, communication partners must agree on a joint protocol to secure their data exchange. The National Institute of Standards (NIST) is in the process of standardizing quantum-safe algorithms ([N23], [CJ+16]). The process consists of several phases: first, submissions of quantum-safe algorithms has been solicited. The submitted algorithms are evaluated, and positively evaluated algorithms become candidates for standards. These candidates are then turned into specifications. The specifications are then reviewed by a global community. Those specifications that did not fail the review become Federal Information Processing Standards (FIPS).

Especially, Kyber and Dilithium have been standardized ([N24a] and [N24b]); note that NIST officially uses the names Module-Lattice-Based Key-Encapsulation Mechanism or ML-KEM, and Module-Lattice-Based Digital Signature or ML-DS in its specifications, which reflects the mathematical underpinnings of the algorithms. Other algorithms are under standardization by NIST also or remain candidates for future standardization. [RB+22] gives an overview of the mathematics of the algorithms finally considered by NIST for standardization.

To support crypto-agility (next section), NIST identified 14 candidates for standardization in a second round [N24c]. These candidates are signature algorithms that will be analyzed according to the phases sketched before.

## 6.5. Crypto-Agility

The algorithms that are currently standardized are believed to be quantum-safe. I.e. there are no known classical algorithms or quantum algorithms that can crack them. But there is no guarantee (in the sense of a mathematical proof) that they cannot be broken in future. In fact, there is some evidence that a couple of them will be finally broken [N21]; and one of the NIST candidates has already been broken [G22]. However, based on the best knowledge of the majority of the cryptographic community the standards currently being rolled out are post-quantum-safe.

Because of the volatility of being post-quantum-safe, any cryptographic infrastructure that is considered to be post-quantum-safe in total must cope with situations in which one or more of the algorithms that are considered to be quantum-resistant today will be cracked. It must be possible to easily extract the corresponding algorithms from the infrastructure such that they will not be used anymore. Furthermore, it should be possible to substitute these algorithms by others that are still considered to be quantum-resistant at this point in time. Thus, the infrastructure must be *crypto-agile* [MH19].

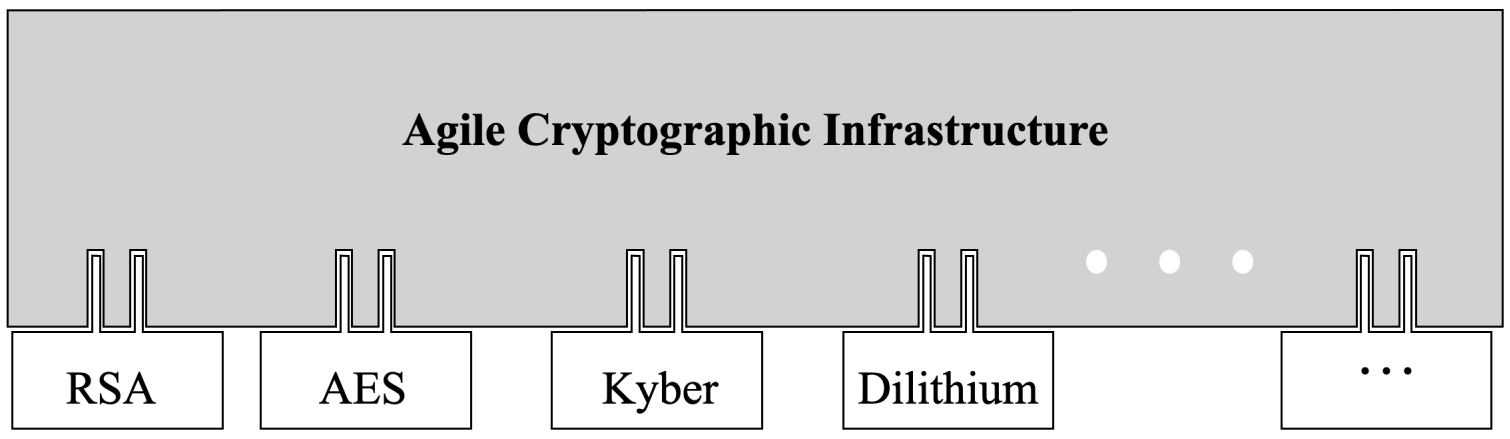

**Fig. 10**. Plugin-Based Cryptographic Infrastructure

Figure 10 depicts the high-level architecture of such an agile cryptographic infrastructure. It is plugin-based, i.e. the implementations of the algorithms are plugins that can be easily exchanged. Also, more plugins can be added whenever implementations of new post-quantum-safe algorithms appear.

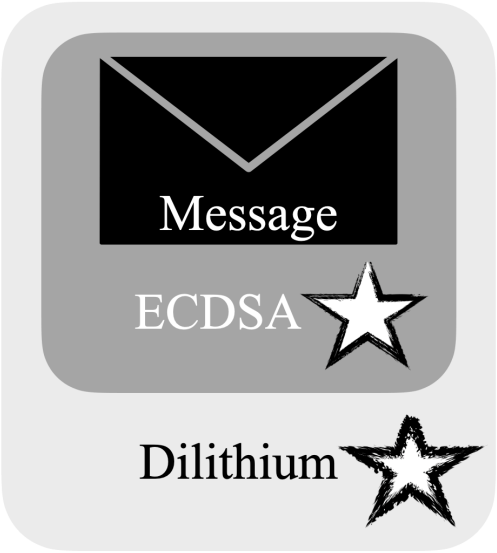

**Fig. 11**. Signing a Message Using the $C \rightarrow Q$ Approach

Furthermore, algorithms can be combined. I.e. a classical algorithm $C$ is not necessarily replaced by its quantum counterpart $Q$, but both algorithms are used together (a.k.a. *hybrid cryptography*). For example, a document may be encrypted based on the classical AES algorithm as well as based on the quantum-safe Kyber algorithm. This is an example of the $C \rightarrow Q$ approach where first the classical algorithm $C$ is applied and next the corresponding quantum algorithm $Q$ is applied on the result of the preceding step. Obviously, $Q \rightarrow C$ is also a valid approach. Classical and quantum algorithms might also be applied in parallel, i.e. a $C \| Q$ approach is followed. For example, a signature may be computed based on a classical algorithm as well as based on a quantum algorithm, and finally both signatures are associated with the signed object. Figure 11 depicts the $C \rightarrow Q$ approach for signing a message: first, the message is signed classically by using Elliptic Curve Digital Signature Algorithm (ECDSA), and the resulting object is signed by means of Dilithium. The advantage of such combined approaches is that they are at least as secure as the current classical algorithm $C$, but it allows to gain trust in the quantum algorithm $Q$ and its implementations.

## 7. In Brief: Sample Open Source Activities

Several of the algorithms that are standardized by NIST (and others) are available as open source implementations. Also, other implementations that use this code in communication protocols, for example, is available in open source too. In this chapter we sketch a particular stream of open source activities having a significant industry support while other activities are not mentioned at all.

### 7.1. OQS

A consortium named Post-Quantum Cryptography Alliance (PQCA) members of which include AWS, CISCO, Google, IBM etc. “aims to support the transition to quantum-resistant cryptography” [PQCA]. Amongst other activities, PQCA runs a project called Open Quantum Safe (OQS) that builds implementations of post-quantum algorithms and usages thereof [OQS].

OQS provides a library called liboqs (see next section) of algorithm implementations. These implementations are also used to build quantum-safe implementations of TLS, SSH etc. [OQSA]. The latter implementations are built by providing branches of OpenSSL, OpenSSH, for example, that make use of liboqs. The quantum-safe branch of TLS, for example, has been used to demonstrate how to enable Apache httpd or Nginx, for example, for becoming quantum-safe.

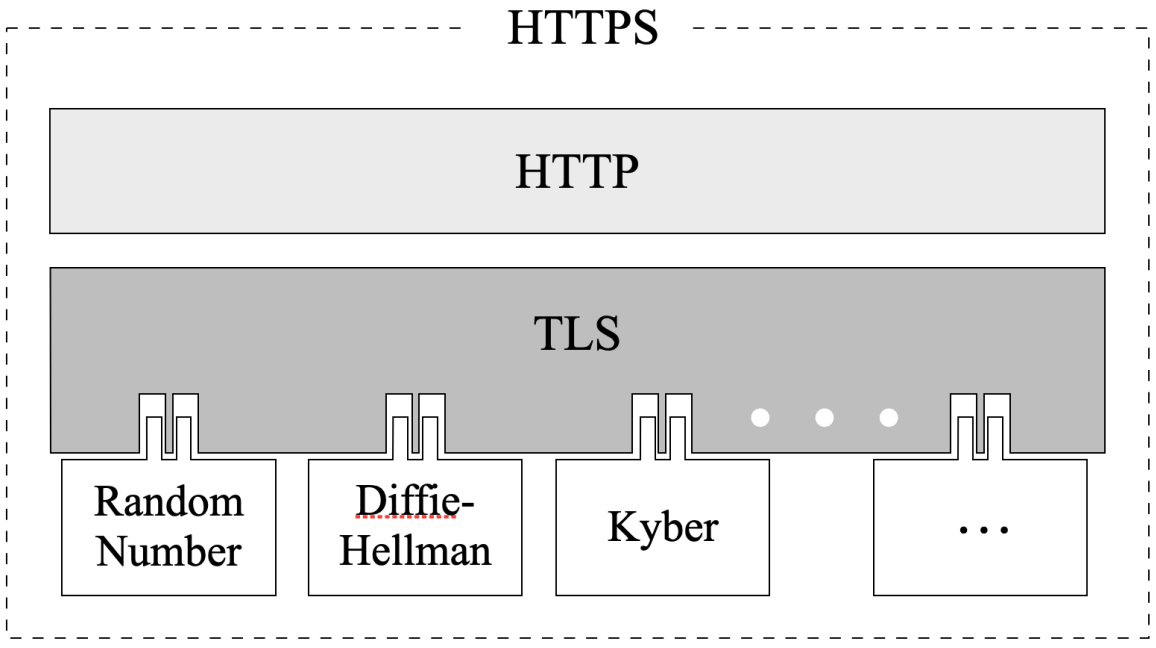

**Fig. 12**. Making HTTPS Quantum-Safe

Figure 12 shows how HTTPS can become quantum-safe: HTTPS is HTTP over TLS. Thus, HTTPS can be turned into a quantum-safe protocol by making TLS quantum-safe. Classically, TLS is using by default an encrypted secret random number-based mechanism or it can be setup to exploit Diffie-Hellman (refer to section 2.6). By using Kyber as key encapsulation mechanism TLS becomes quantum-safe and this implies the quantum-safeness of HTTPS too.

### 7.2. liboqs

A collection of implementations of quantum-safe algorithms is provided by OQS under MIT license as an open source C-library on Github [LOQS]. Wrappers for other programming languages are provided, for example for Java, Python, C++. Beside implementations of Kyber and Dilithium (or ML-KEM and ML-DS, respectively) the library contains also implementations of other algorithms for key encapsulation mechanisms and signature schemes.

It is explicitly mentioned on the websites of liboqs that the library is of prototypical nature only, i.e. it should not be used in production environments and it should not be used to protect sensitive data. In contrast, the goal of the Post-Quantum Cryptography Coalition PQCC [PQCC] (whose founding members include members of the PQCA) lists explicitly the creation of corresponding production-quality code, and ensuring cryptographic agility (refer to section 6.5).

### 7.3. Sandwich

Sandwich is an open source implementation that provides cryptographic agility enabling post-quantum-safeness by using liboqs [S]. It defines a unified API for choosing cryptographic algorithms and protocols at runtime. Thus, no code must be modified to exchange algorithm implementations (but proper configuration files must be passed at runtime).

## 8. Conclusion

In this paper, we described the origins of the need for efforts subsumed by the term post-quantum security. The relevant number theoretic background has been provided as the underpinning of security schemes based on factorization. The algebra of elliptic curves has been sketched and security schemes on top of it was described. We revealed that both schemes are in fact relying on the hardness of computing discrete logarithms. Shor's algorithm was explained in brief, showing that security schemes relying on discrete logarithms are broken as soon as powerful enough quantum computers become available (which may very well be within the next couple of years).

Lattice-based cryptography is seen as a rescue. We introduced lattices and fundamentally hard problems that provide the base for corresponding security schemes. To avoid huge keys that are implied by high security levels required in practice, algebraic lattices have been introduced and two corresponding cryptographic schemes called Kyber and Dilithium have been discussed.

Standardization efforts in this domain have been reported, and related open source activities were mentioned. The "harvest now, decrypt later" attack was described and impact on organizations revealed. Crypto-agility as an important principle has been emphasized.

**Author Contributions**: Writing – original draft, F.L.; Writing – review & editing, J.B. All authors have read and agreed to the published version of the manuscript.